\documentclass[journal]{IEEEtran}
\usepackage{amsmath,amssymb,amsfonts,amsthm,mathtools}
\usepackage{graphicx}
\usepackage{textcomp}
\usepackage{url}
\usepackage{cases}
\usepackage[utf8]{inputenc}

\usepackage[linesnumbered,ruled,vlined]{algorithm2e}
\usepackage{algpseudocode}

\DeclarePairedDelimiter{\ceil}{\lceil}{\rceil}

\usepackage{multirow}
\usepackage{dcolumn,booktabs}

\usepackage[caption=false]{subfig} 
\makeatletter
\renewcommand{\fnum@figure}{Fig. \thefigure}
\makeatother

\usepackage{hyperref}
\hypersetup{colorlinks=true,linkcolor=blue,urlcolor=blue}

\usepackage[normalem]{ulem}

\usepackage[noadjust]{cite}

\begin{document}

\title{User Clustering in mmWave-NOMA Systems with User Decoding Capability Constraints for B5G Networks}

\author{Aditya S. Rajasekaran~\IEEEmembership{(Member,~IEEE)}, Omar~Maraqa, Hamza~U.~Sokun, Halim Yanikomeroglu~\IEEEmembership{(Fellow,~IEEE)}, and Saad Al-Ahmadi%

\thanks{The work of A. S. Rajasekaran was supported by Ericsson Canada Inc. in the form of tuition assistance. This work was also supported in part by a Discovery Grant of the Natural Sciences and Engineering Research Council of Canada and in part by the King Fahd University of Petroleum \& Minerals (KFUPM) through grant number SB191049. Corresponding author: Aditya~S.~Rajasekaran (aditya.sriram.rajasekaran@ericsson.com)}%
\thanks{A.~S.~Rajasekaran is with the Department of Systems and Computer Engineering, Carleton University, Ottawa, ON K1S 5B6, Canada and with Ericsson Canada Inc, Ottawa, ON K2K 2V6, Canada (email: aditya.sriram.rajasekaran@ericsson.com).}%
\thanks{O.~Maraqa and S. Al-Ahmadi are with the Department of Electrical Engineering, King Fahd University of Petroleum \& Minerals, Dhahran-31261, Saudi Arabia (e-mail: g201307310@kfupm.edu.sa; saadbd@kfupm.edu.sa).}%
\thanks{H.~U.~Sokun is with Ericsson Canada Inc, Ottawa, ON K2K 2V6, Canada (email: hamza.sokun@ericsson.com).}%
\thanks{H. Yanikomeroglu is with the Department of Systems and Computer Engineering, Carleton University, Ottawa, ON K1S 5B6, Canada (email: halim@sce.carleton.ca).}%

\thanks{$\copyright$ 2020 IEEE. Personal use of this material is permitted. Permission from IEEE must be obtained for all other uses, in any current or future media, including reprinting/republishing this material for advertising or promotional purposes, creating new collective works, for resale or redistribution to servers or lists, or reuse of any copyrighted component of this work in other works.}%
}

\markboth{Accepted by IEEE Access, November 2020}%
{Rajasekaran~\MakeLowercase{\textit{et al.}}: User Clustering in mmWave-NOMA Systems with User Decoding Capability Constraints for B5G Networks}

\maketitle

\begin{abstract}
\textcolor{black}{This paper proposes a millimeter wave-NOMA (mmWave-NOMA) system that takes into account the end-user signal processing capabilities, an important practical consideration. The implementation of NOMA in the downlink (DL) direction requires successive interference cancellation (SIC) to be performed at the user terminals, which comes at the cost of additional complexity. In NOMA, the weakest user only has to decode its own signal, while the strongest user has to decode the signals of all other users in the SIC procedure. Hence, the additional implementation complexity required of the user to perform SIC for DL NOMA depends on its position in the SIC decoding order. Beyond fifth-generation (B5G) communication systems are expected to support a wide variety of end-user devices, each with their own processing capabilities. We envision a system where users report their SIC decoding capability to the base station (BS), i.e., the number of other users signals a user is capable of decoding in the SIC procedure. We investigate the rate maximization problem in such a system, by breaking it down into a user clustering and ordering problem (UCOP), followed by a power allocation problem. We propose a NOMA-minimum exact cover (NOMA-MEC) heuristic algorithm that converts the UCOP into a cluster minimization problem from a derived set of valid cluster combinations after factoring in the SIC decoding capability. The complexity of NOMA-MEC is analyzed for various algorithm and system parameters. For a homogeneous system of users that all have the same decoding capabilities, we show that this equates to a simple maximum number of users per cluster constraint and propose a lower complexity NOMA-best beam (NOMA-BB) algorithm. Simulation results demonstrate the performance superiority in terms of sum rate compared to orthogonal multiple access (OMA) and traditional NOMA clustering schemes that do not incorporate individual users' SIC decoding capability constraints.}
\end{abstract}


\section{Introduction}\label{sec:intro}

\IEEEPARstart{B}eyond fifth-generation (B5G) communication systems are expected to support a large number of connected users at a time, each with different processing capabilities and requirements. \textcolor{black}{The massive machine-type connectivity (mMTC), also called the Internet of Things (IoT), as well as the ultra-reliable low latency communication (URLLC) use-cases, are expected to bring many different types of connected users into the system compared to traditional mobile broadband users \cite{ghosh2019B5G}. Thus, B5G systems need to support a very large number of low-cost devices for the IoT connections in addition to the traditional high data rate mobile broadband connections that are also growing exponentially. The latest Ericsson mobility report \cite{ericssonMobilityReport} estimates upwards of 30 billion connected users by 2023, with more than 50\% coming from IoT connections. This puts enormous spectral efficiency requirements on the B5G wireless communication systems}.

The mmWave spectrum offers a large amount of bandwidth to scale up the capacity from the cellular networks that operate today in the sub-6 GHz range. Further, non-orthogonal multiple access (NOMA) techniques offer a way to serve multiple users in the same orthogonal resource, e.g., time, frequency, orthogonal frequency division multiplexing (OFDM) resource block (RB), etc., by separating the users in the power domain instead (PD-NOMA). Hence, when combined, mmWave-NOMA has the potential to serve the high rates and massive connectivity demands of B5G networks. Additionally, the high level of correlation amongst users channels in mmWave, makes them ideal for the formation of user clusters to be served by a single beam and separated in the power domain through NOMA \cite{cui2018OptimizationBased, zhu2019mmWaveNOMA, cui2018unsupervised}. 

The survey in \cite{maraqa_nomasurvey2020} shows that the key aspects of achieving a good performance in NOMA systems are user clustering, user ordering, beamforming, and power allocation. User clustering refers to the selection of users to serve in a NOMA cluster, typically in a beam via beamforming techniques. User ordering refers to the order in which successive interference cancellation (SIC) is applied at the users in the downlink. Power allocation techniques are then used to allocate the right amount of power to each user in the cluster, so that SIC decoding can be successful and each users target rates are met. The focus of this paper is user clustering and user ordering. 

As we group users in NOMA clusters, the weakest user only has to decode its own signal, while the strongest user has to decode the signals of all other users in the SIC procedure. The decoding of other users' signals requires significant additional processing capability, in terms of hardware capability, energy consumption, etc.~\cite{tse2005fundamentals, tabassum2017Challenges}. The authors in \cite{islam2017survey_SICdecodingComplexity} identified this SIC decoding complexity as the first major practical implementation issue for NOMA. \textcolor{black}{NOMA is expected to support a wide variety of end-user devices in B5G systems, each with different signal processing capabilities~\cite{islam2017survey_SICdecodingComplexity, saad20206G}}. Hence, each user has its own limitations on the number of other users signals that it can decode. \textcolor{black}{We term this the SIC decoding capability of the user. For NOMA, this SIC decoding capability translates to the number of other users signals a DL user can decode before decoding its own signal. This SIC decoding capability can be easily communicated to the BS during connection setup. For IoT devices, this could be as low as zero or one, while for high-end smartphones this can be a much higher value due to the differences in hardware processing capability. Hence, when implementing NOMA in the DL, the BS needs to respect this SIC decoding capability limit of the user when it orders users to be served in a NOMA cluster as we discuss further in the motivation in Section \ref{sec:motivation}. }


\subsection{Related Work}


In \cite{ali2019clustering}, the authors highlight the tight coupling between user clustering, cluster sizes, and user ordering on the performance of NOMA systems. In typical NOMA works from the literature, the user pairing or user clustering schemes have been designed to group two users per cluster \cite{hao2017LettermmWaveUP, chen2018LetterUserPairing}, or some fixed number of users per cluster \cite{ali2019NOMAclustersizes}, respectively. In \cite{ali2019NOMAclustersizes}, the optimum cluster sizes from a performance perspective is analyzed. However, in this paper, we focus on the cluster size as a constraint. More importantly, it is not just a generic constraint that limits the number of users in the cluster, but there is a constraint from each user in the cluster on how many other users signals it can decode in the SIC decoding order. When it comes to the SIC decoding order within a cluster, as highlighted in \cite{ding2020unveiling}, users are typically ordered either based on their effective channel gains, \textcolor{black}{i.e., channel gains after considering the beamforming weights}, or based on their quality of service (QoS) using a cognitive radio concept. In this paper, we focus on the effective channel gain strategy as we assume all users have the same QoS.

\textcolor{black}{Unlike multi-user MIMO (MU-MIMO), where correlated users are difficult to separate by individual  beams, such correlated users can easily be grouped together in a NOMA cluster~\cite{maraqa_nomasurvey2020}. In mmWave systems, the users' channels are highly correlated due to the highly directional nature of mmWave transmission~\cite{xiao2018Joint, Ding2017BF_mmWaveNOMA}. The user clustering schemes in mmWave-NOMA systems typically exploit the high correlation amongst users channels to cluster correlated users together, e.g., \cite{cui2018OptimizationBased, wangCuiDing2019_stackleberg}.  In~\cite{shao2020angle}, the authors use an angle-domain NOMA scheme that schedules one cell-center and one cell-edge user in a NOMA pair, for each beam in each cell.  Recent works in mmWave-NOMA systems have also used machine learning clustering techniques to identify correlated users and group them in NOMA clusters \cite{cui2018unsupervised, Marasinghe2020AHC, ren2019EMclustering}. Further, in mmWave systems, since it is often infeasible to scale up the number of transceivers with the number of antennas, studies in mmWave-NOMA systems often use either analog beamforming (BF) with a single RF chain~\cite{xiao2019user, uwaechia2020spectrum, xiao2018Joint} or a hybrid BF design with a reduced number of RF chains~\cite{dai2019Hybrid, fan2019channel}.}

\subsection{Motivation and Contributions} \label{sec:motivation}

\textcolor{black}{In practical deployments, the typical clustering approaches described above in the related work in mmWave-NOMA have two important limitations.} First, they can lead to arbitrarily large and uneven cluster sizes. If we have a system model where one cluster is served on one channel, this could lead to over-use on one channel and under-use on other channels. More than the imbalance in resource usage, large cluster sizes mean the users at the end of the SIC decoding orders need to decode a very large number of users. This is particularly an issue in dense deployments, where large clusters of correlated users can exist. The second important limitation with these algorithms is there is no flexibility incorporated to account for the SIC decoding capability limitations of each individual user. Concretely, just finding groups of correlated users, can lead to cluster formations where the individual user decoding capability limits of some users are not respected, i.e., users are placed in SIC decoding positions in a cluster that require them to decode the signals of \textcolor{black}{a greater} number of users than their \textcolor{black}{indicated SIC decoding capability}.

\textcolor{black}{The} clustering schemes from the mmWave-NOMA literature that focus on finding correlated users, e.g., \cite{cui2018unsupervised}, can be modified to arbitrarily divide the groups of correlated users the algorithm identified into different clusters, served on different channels. Users then need to be decoded in the order of their decoding capabilities, rather than the effective channel gains. Even when users are decoded in the order of SIC decoding capability, further orthogonal channels might be needed if some users' constraints are not met. All such workarounds to meet the practical SIC decoding capability constraints of users in real deployments would erode the gains from the clustering algorithms that strived to find good sets of correlated users, with sufficient separation between the clusters. Instead, if the clustering algorithm was able to consider the user decoding capability requirements as part of its input, it would be better able to construct clusters that maximize the overall spectral efficiency, while taking into account these individual user decoding capability constraints. This is the motivation for the work presented in this paper.

Against this background, in this paper, we \textcolor{black}{investigate a rate maximization problem for a mmWave-NOMA system that takes into consideration the SIC decoding capability of each individual user in the system. We break down the problem into a user clustering and ordering problem (UCOP), which is the focus of this paper, followed by a power allocation (PA) problem.} We consider a single-cell mmWave-NOMA equipped base station (BS) that applies analog beamforming (ABF) in a fixed set of directions uniformly distributed around the cell coverage area. \textcolor{black}{A NOMA cluster of users will be served on an orthogonal channel, e.g., a time channel or OFDM resource block,  using  one  of  these pre-defined beams. In this way, the UCOP can be framed as a cluster minimization problem in order to minimize the number of orthogonal channels used to serve the required number of users, while respecting each individual user's SIC decoding capability constraints. We propose two algorithms to solve this UCOP. The first one we term NOMA-minimum exact cover (NOMA-MEC), as we decompose the problem into a MEC problem, a known NP-complete problem~\cite{MEC}. For a homogeneous system where all users have the same SIC decoding capability, we propose a less complex NOMA-best beam (NOMA-BB) algorithm. The key aspects of the two algorithms are outlined next. }

\textcolor{black}{In both algorithms, the BS uses the cosine similarity metric that aligns a user's channel with the set of possible beam directions to rank the best beams for each user. The BS then chooses the best beams to form a candidate beam list for each user, with the number of beams in this list a configurable parameter that can be tuned for a complexity-performance trade-off as we discuss in-depth in this paper. This step identifies users that can potentially cluster with each other. User ordering in any cluster is done in the order of the effective channel gains. NOMA-MEC then takes the SIC decoding capability of the users into account, and builds a list of valid cluster combinations such that the SIC decoding is done in the order of the users' channel gains and each user SIC decoding capability is respected. Using this candidate list, NOMA-MEC is able to frame the problem as a MEC problem where the goal is to serve all the users in the least number of channels from the designed set of valid cluster combinations. }

In a homogeneous system, the user decoding capability constraints of the users translate to limiting the number of users per cluster, as any user ordering within that cluster will satisfy each user's decoding constraints since they are all equal. Such a homogeneous system with a restriction on the maximum number of users per cluster is what is typically considered in user clustering algorithms in the literature, e.g., \cite{ali2019NOMAclustersizes}. In our case, the homogeneous system is just a special case of the heterogeneous system where all users have the same SIC decoding capabilities, and so the NOMA-MEC algorithm can still be run. However, for this simpler homogeneous system, we also propose a simpler NOMA-best beam (NOMA-BB) algorithm that demonstrates comparable performance to NOMA-MEC when we have the special setting of all users having the same decoding capability. Finally, we demonstrate the performance superiority of NOMA-MEC compared to orthogonal multiple access (OMA) as well as the additional flexibility the NOMA-MEC scheme offers in heterogeneous systems compared to other NOMA clustering schemes like NOMA-BB that target a fixed number of users per cluster.

\begin{table*}[!htb]
\centering
\caption{List of parameters.}
\label{Table: List of parameters}
\begin{tabular}{|c|l|}
\hline
\textbf{Notation} & \textbf{Description} \\ \hline


$\boldsymbol{h}_{u}$, $\alpha_u$, $L$, $r_{u}^{\eta}$ &mmWave channel parameters\\ \hline
$\boldsymbol{a}(\theta_{u})$, $\phi_u$, $D$, $\lambda$ &Antenna array related parameters \\ \hline
$M$ &Number of antennas at BS \\ \hline
$N$ &Total number of users in the system \\\hline
$K$ &Number of clusters \\\hline

$C = \{C_1, .., C_K\}$ & The set of $K$ clusters\\\hline
($B, \bar\theta$) & \begin{tabular}[l]{@{}l@{}} \textcolor{black}{Coverage area $\bar\theta$ is divided into $B$ parts, which}\\\textcolor{black}{the BS uses to generate $B+1$ candidate beams}\end{tabular} \\\hline

$B_c = \{\textnormal{Beam-}0,.., \textnormal{Beam-}B\}$ &The set of $B+1$ candidate beams\\\hline
$W_c = \{\boldsymbol{w}_0,.., \boldsymbol{w}_B\}$ &The set of $B+1$ candidate precoding vectors \\\hline
$B_u, b$ &User-beam set for user-$u$ containing $b$ beams \\\hline
\textcolor{black}{$B_u^*$} & Best beam in $B_u$, i.e., $B_u$ with $b=1$ \\\hline
($C_k$, \textcolor{black}{$N_{C_k}$}, $b_k$, $\boldsymbol{w}_{b_k}$) & \begin{tabular}[l]{@{}l@{}}The cluster $C_k$, containing \textcolor{black}{$N_{C_k}$}  users, served\\on beam-$b_k$ with a precoding vector $\boldsymbol{w}_{b_k}$ \end{tabular}\\\hline
$s_{k}$ &The transmitted signal to cluster-$k$\\\hline
$y_{u}$ &The received signal at user-$u$\\\hline
$p_u$ &The power allocated to user-$u$\\\hline
$P$ &The power available per channel\\\hline
\textcolor{black}{$\xi_u$} &The noise at user-$u$\\\hline
$\sigma^2$ &The noise power\\\hline
$\pi_{k} (j)$ &The user index for the $j$-th decoded user in the $k$-th cluster\\\hline

\textcolor{black}{$\Gamma_{\pi_k(j^{\prime})}^{\pi_k(j)}$} &\textcolor{black}{The SINR when decoding user $\pi_k(j)$ at user $\pi_k(j^{\prime})$, where $j^{\prime} > j$} \\\hline

\textcolor{black}{$\Gamma_{\pi_k(u)}^{\pi_k(u)}$} &\textcolor{black}{The SINR when decoding user $\pi_k(u)$ own signal}\\\hline

$R_{\textnormal{sum}}$ / $R_{\textnormal{OMA}}$ &\begin{tabular}[l]{@{}l@{}}The effective sum rate of the system\\that adopted NOMA/OMA scheme\end{tabular} \\\hline
$R_{k}$ &The sum rate of the users within cluster-$k$\\\hline
\textcolor{black}{$\Gamma_{\textnormal{min}}$} &Users minimum QoS SINR\\\hline
$d_{\textnormal{max}}$ & \begin{tabular}[l]{@{}l@{}} The maximum decoding capability among the\\  $N$ users in the system\end{tabular}\\\hline
$d_u$ &Decoding capability of user-$u$, in the range $[0,d_{\textnormal{max}}]$\\\hline
$m$ &A NOMA-BB algorithm parameter between 1 and $d_{\textnormal{max}}$\\\hline
$d_{\pi_k(j)}$ & \begin{tabular}[l]{@{}l@{}}The decoding capability of the $j$-th\\decoded user in the $k$-th cluster \end{tabular}\\\hline
$C_v$ &A list of viable candidate cluster options \\\hline
$n_b$ & \begin{tabular}[l]{@{}l@{}} The number of users that have beam-$b$\\in their user-beam set \end{tabular}\\\hline
$b_{\textnormal{th}}$ & \begin{tabular}[l]{@{}l@{}} User is in the coverage area of $b_{th}$ beams from the\\ list of candidate beams in $B_c$ \end{tabular}\\\hline
$I_1$&Number of iterations examined in step-1 of NOMA-MEC \\\hline
$I_2$& \begin{tabular}[l]{@{}l@{}}Number of possible combinations for step-2 of\\ NOMA-MEC to examine, i.e., number of elements in $C_v$\end{tabular}\\\hline
$u_{i,k}$, $z_{u,c}$, $x_c$ &Binary variables defined as part of NOMA-MEC algorithm\\\hline
$l,c_s, C_t,C_\textnormal{sorted}$ &Variables defined as part of NOMA-MEC algorithm\\\hline
$(.)^T$, $(.)^H$ &Transpose and Hermitian transpose\\ \hline
$\ceil{.}$ &Ceiling function\\\hline
\end{tabular}%
\vspace{-1em}
\end{table*}

Our contributions can thus be summarized as follows:
\begin{itemize}
    \item We design a joint user-clustering, user ordering, and beamforming scheme in a mmWave-NOMA ABF system with a fixed set of candidate beams that minimizes the number of clusters required to serve all the users, subject to each individual users decoding capability and beamforming constraints. Each NOMA cluster is served on one orthogonal channel, so minimizing the number of clusters also minimizes the number of channel uses. \textcolor{black}{Together with a power allocation scheme per cluster, we maximize the sum rate of the system.} 
    \item To the best of the authors' knowledge, this is the first NOMA work that considers the individual SIC decoding capability of each user when doing NOMA clustering and ordering. The proposed scheme is ideally suited for a mmWave-NOMA deployment involving a low-cost small-cell BS with only one RF chain, supporting a large number and variety of connected users, from low-cost IoT devices with limited processing capabilities to high-end smartphones with much larger processing capabilities. From the perspective of NOMA in the downlink, the processing capability of the user primarily impacts the SIC decoding capability, i.e., the number of other users signals a user can decode every channel use.
    \end{itemize}
    
\subsection{Paper Organization}

The rest of this paper is organized as follows. In Section \ref{sec:sysModel}, the system model and problem formulation that aims to minimize the number of clusters required to serve the given users are presented. In Section \ref{sec:algo}, we detail the proposed NOMA-MEC and NOMA-BB algorithms. Detailed simulation results for the proposed algorithms, including a complexity analysis for different algorithm parameters, are presented in Section \ref{sec:simulation}. Finally, concluding remarks are provided in Section \ref{sec:conclusion}. Table~\ref{Table: List of parameters} lists the notations used in this paper.

\section{System Model and Problem Formulation}\label{sec:sysModel}



Consider a mmWave-NOMA single-cell BS equipped with $M$ antennas serving $N$ single-antenna users, each with a minimum QoS constraint. We use the single path mmWave channel model used in several mmWave-NOMA papers \cite{cui2018unsupervised, cui2018OptimizationBased, Ding2017BF_mmWaveNOMA} to model the mmWave channel between the BS and user-$u$ as follows:

\begin{equation} \label{eq: channel response}
\boldsymbol{h}_{u} =\boldsymbol{a}(\theta_{u})\frac{\alpha_{u}}{\sqrt{L}  \big(1 + r_{u}^{\eta} \big)},
\end{equation}

\noindent where $L$ denotes the number of paths, $r_{u}$ denotes the distance between the BS and user-$u$, $\eta$ denotes the path loss exponent and $\alpha_u$ denotes the complex channel gain for user-$u$. The parameter $\theta_u$ represents the physical angle of departure and for a uniform linear array (ULA), the normalized angle is defined as $\phi_u = 2\frac{D}{\lambda} \sin(\theta_u)$, where $D$ is the separation between elements of the antenna and $\lambda$ is the wavelength of the carrier signal~\cite{Ding2017BF_mmWaveNOMA}. The term $\boldsymbol{a}(\theta_{u})$ represents the steering vector and for a ULA can be represented as 

\begin{equation} \label{eq: steering vector}
\begin{split}
\boldsymbol{a}(\theta_{u}) &=[1, e^{-j2\pi  \frac{D}{\lambda} \sin(\theta_{u})}, .., e^{-j2\pi (M-1) \frac{D}{\lambda} \sin(\theta_{u})}]^{T} \\
& = [1, e^{-j\pi  \phi_{u}}, .., e^{-j\pi (M-1) \phi_{u}}]^{T}.
\end{split}
\end{equation}

\begin{figure}[t!]
\centering
\vspace{-1em}
\includegraphics[width=0.48\textwidth]{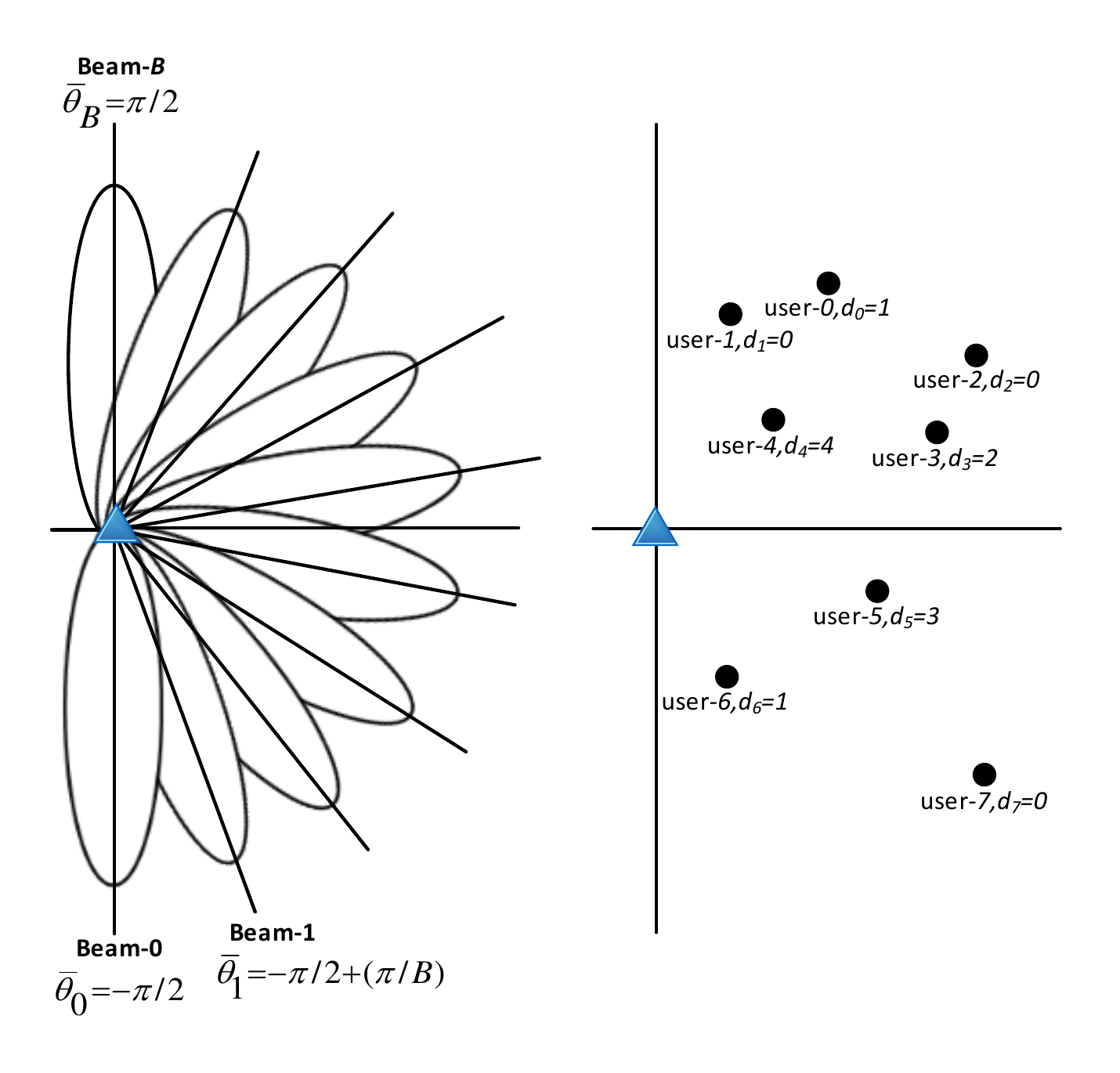}
\caption{The ABF scheme used by the BS is illustrated on the left with a fixed set of $B+1$ precoding weights, creating $B+1$ candidate beam directions to choose from when serving a NOMA cluster of users in one orthogonal channel. On the right, the $N$ users, each with their own SIC decoding capability, is illustrated.} 
\label{fig:systemModel}
\vspace{-1em}
\end{figure}


\textcolor{black}{Analog beamforming is used since} only one radio frequency (RF) chain is available at the BS, \textcolor{black}{typical of small-cell deployments where low hardware cost and power consumption is essential, e.g., \cite{xiao2019user, Ding2017BF_mmWaveNOMA}}. Hence, only one beam can be transmitted at a time, which we equate to forming one beam to serve one cluster of NOMA users per channel use. \textcolor{black}{Since we use ABF that can only generate one beam at a time, we use a time-division strategy to alternate between the different clusters.} 

As the left part of Fig. \ref{fig:systemModel} illustrates, the entire coverage region, $\bar\theta$, from $-\pi/2$ to $\pi/2$ is covered by a set of $B+1$ candidate beams, with significant overlap between the candidate beams. A NOMA cluster of users will be served on an orthogonal channel using one of these candidate beams. Each beam-$b$ in this candidate list has the following precoding vector,
\begin{equation}\label{eq:candidateBeams}
\begin{split}
\boldsymbol{w}_b &= \boldsymbol{a}(\bar\theta_b), \forall b \in [0,B]\\
\end{split}
\end{equation}

\noindent where the parameter $\bar\theta_b$ is
\begin{equation}\label{eq:candidateBeams-b}
\begin{split}
\bar\theta_b = -\pi/2 + (b \times \pi/B). \\
\end{split}
\end{equation}

In this way, we uniformly divide this entire coverage region, $\bar\theta$, into $B$ equal angles, effectively forming a set of $B+1$ candidate beams, as illustrated in the left part of Fig. \ref{fig:systemModel}. The $B+1$ beams can be thought of as a choice of $B+1$ different precoding vectors based on \eqref{eq:candidateBeams}, such that collectively, the steering vectors of the $B$ candidate precoding vectors uniformly cover the entire region of $\bar\theta = -\pi/2$ to $\pi/2$ or $\bar\phi = -1$ to $1$. We let $B_c$ represent this list of candidate beams, such that $B_c = \{\textnormal{Beam-}0,.., \textnormal{Beam-}B\}$, with their respective list of candidate precoding vectors being $W_c = \{\boldsymbol{w}_0,.., \boldsymbol{w}_B\}$, as illustrated in Fig.\ref{fig:systemModel}. 

A NOMA cluster of users will be served on an orthogonal channel using one of the $B+1$ precoding vectors in the candidate list. \textcolor{black}{In our system model, the orthogonal channel is a time slice, hence each cluster will be served in one time slice}. A total of $K$ such clusters are formed to serve the $N$ users. This equates to requiring $K$ channel uses, \textcolor{black}{i.e., $K$ time-slices}, to serve the $N$ users, $K \leq N$. \textcolor{black}{Let $C = \{C_1, .., C_K\}$ represent the $K$ clusters required to serve the $N$ users, where $C_k$ refers to the $N_{C_k}$ users, $N_{C_k} \leq N$, selected to serve in the cluster with index-$k$. For each cluster $C_k$, a beam-$b_k$ with a precoding vector $\boldsymbol{w}_{b_k}$ is selected from the set of $B+1$ possible beam options in $B_c$.}  





We exploit the high correlation in the mmWave channels as follows. We assume the BS has access to the full channel state information (CSI) vector of each user, $\boldsymbol{h}_{u}$, from \eqref{eq: channel response} \cite{cui2018unsupervised, zhu2019mmWaveNOMA}. Additionally, the BS knows the precoding vectors of each beam, $\boldsymbol{w}_b$, in the candidate set. The BS can use the cosine similarity metric between the user's channel vector and the precoding vector of each beam to determine the level of correlation between the user and the beam. This metric has been used in several mmWave-NOMA works for user clustering to determine the correlation between users in \cite{cui2018unsupervised, Marasinghe2020AHC}, and between users and random beams in \cite{Ding2017BF_mmWaveNOMA}. Using similar steps as these works, we can derive the cosine similarity between a user-$u$ with channel $\boldsymbol{h}_{u}$ and a beam-$b$ with precoding vector $\boldsymbol{w}_{b}$ here as follows:


\begin{equation} \label{eq:cosS}
\begin{split}
\cos(\boldsymbol{h}_{u},\boldsymbol{w}_{b}) & =\frac{\mid \boldsymbol{a}(\phi_{u})^{H} \boldsymbol{a}(\phi_{b}) \mid}{M}\\
      & = \frac{\mid \sum_{i=0}^{M-1} e^{-j\pi i (\phi_u - \phi_b)}\mid }{M}\\
      & = \mid\frac{\sin \big( \frac{\pi M(\phi_u - \phi_b)}{2} \big) }{M \sin \big(  \frac{\pi(\phi_u - \phi_b)}{2} \big) }\mid\\
      & = F_M  \big(\pi [\phi_u - \phi_b]\big) ,
\end{split}
\end{equation}

\noindent where $\phi_u$ and $\phi_b$ are the normalized directions of the user and beam respectively, $H$ represents the Hermitian transpose and $F_M$ represents the Fejer Kernel, whose properties dictate that as $\mid \phi_u - \phi_b \mid$ increases, $\cos(\boldsymbol{h}_{u},\boldsymbol{w}_{b})\rightarrow 0$. In other words, if the beam and users directions are well aligned, the cosine similarity metric is high and it reflects that it is suitable to schedule the user on a cluster served by the beam-$b$. In this way, the BS builds a user-beam set, $B_{u}$, for each user-$u$. This user-beam set, $B_u$ consists of $b$ beams each, by selecting the best $b$ beams for each user using the cosine similarity metric from \eqref{eq:cosS}. The parameter $b$ is a tunable parameter, as we discuss later in Section \ref{sec:algo}. We note that based on the choice of $M$ and $B$, the beams are highly overlapping in nature and so a single user can be served by more than one beam, while still benefiting from a good beamforming gain. Alternatively, the user can select its best beams using the typical ABF approach for the mmWave in new radio (NR) standard \cite{ABF_NR_VTC2018}, but that is beyond the scope of this paper and is a topic for future work. It is also worth mentioning that our BF scheme is different from the random BF scheme in \cite{Ding2017BF_mmWaveNOMA}, where a random beam is generated with precoding vector $\boldsymbol{w}=\boldsymbol{a}(\theta)$, $\theta \in [-\pi/2, \pi/2]$ and all users with a high cosine similarity with that beam are then scheduled. In our scheme, while also ABF with a similar precoding vector, we are not randomly generating the beams, but instead selectively choosing an appropriate beam for a NOMA cluster from a given set of candidate options, $B_c$. 
 
 For cluster $C_k$, the BS applies superposition coding (SC) for the selected \textcolor{black}{$N_{C_k}$} users as follows:
\begin{equation} \label{eq: transmitted signal}
s_{k} = \sum_{u=1}^{N_{C_k}} \sqrt{p_u} s_{k,u},
\end{equation}

\noindent where $p_u$ represents the power allocated to user-$u$ with \textcolor{black}{$\sum_{u=1}^{N_{C_k}} p_u \leq P$}, where $P$ denotes the power available to the BS per-channel use. The received signal at user-$u$ \textcolor{black}{in cluster $C_k$} is 


\begin{equation}
\begin{split}
y_{u} & = \boldsymbol{h}_{u}^{H}\boldsymbol{w}_{b_k}s_{k} + \xi_{u},\\
      & = \underbrace{\boldsymbol{h}_{u}^{H}\boldsymbol{w}_{b_k} \sqrt{p_u} s_{k,u}}_{\textnormal{Desired signal}} + \underbrace{\boldsymbol{h}_{u}^{H}\boldsymbol{w}_{b_k} \sum_{u \neq v, v=1}^{n_{k}}\sqrt{p_v}s_{k,v}}_{\textnormal{Inter-user interference}} + \underbrace{\xi_{u}}_{\textnormal{Noise}}.
\end{split}
\end{equation}


In the SIC procedure, let $\pi_{k} (j)$ denote the user index for the $j$-th decoded user in the cluster $C_{k}$ serving \textcolor{black}{$N_{C_k}$ users, $j \leq N_{C_k}$}. This $j$-th user then needs to decode and subtract all the messages for all users $\{\pi_k(1),..,\pi_k(j)\}$.  The signal-to-interference-plus-noise ratio (SINR) when decoding user $\pi_k(j)$ at user $\pi_k(j^{\prime})$, $j^{\prime} > j$ can be represented as

\begin{equation} \label{eq: SINR of weak at the strong}
 \Gamma_{\pi_k(j^{\prime})}^{\pi_k(j)}= \frac{p_j |\boldsymbol{h}_{(j^{\prime})}^{H}  \boldsymbol{w}_{b_k}|^2}{|\boldsymbol{h}_{(j^{\prime})}^{H}  \boldsymbol{w}_{b_k}|^2 \sum_{v>j}^{N_{C_k}} p_v +  \sigma^2},
\end{equation}

\noindent where $\sigma^2$ represents the noise power. Let $R_k$ denote the rate achieved in NOMA cluster $C_k$. The effective sum rate of the system, $R_{\textnormal{sum}}$ can then be expressed as the sum of the rates, $R_{k}$, achieved in each of the $K$ clusters over which all $N$ users are served divided by the number of clusters, since each cluster is served by one channel. The effective sum rate can thus be represented as
\begin{equation} \label{eq: Sum rate}
\begin{split}
 R_{\textnormal{sum}} & =\frac{\sum_{k=1}^{K} R_k}{K}\\
      & = \frac{\sum_{k=1}^{K} \sum_{u \in C_k} \log_{2}  \big(1 +  \Gamma_{\pi_k(u)}^{\pi_k(u)} \big)}{K},
\end{split}
\end{equation}


\noindent expressed in \textcolor{black}{bits per second (bps) per channel-use and the term $\Gamma_{\pi_k(u)}^{\pi_k(u)}$ refers to the SINR when decoding the $u$-th user's own signal in the SIC decoding procedure}. For OMA, where each user has to be served in an individual channel, $K$ channels are required to serve the $N$ users, i.e., $K=N$. Each user will be served in its best beam from $B_u$ with a precoding vector $\boldsymbol{w}_u$. This gives us an effective sum rate of 
\begin{equation}
    R_{\textnormal{OMA}} = \frac{\sum_{u=1}^{N} \log_{2}(1 + \frac{P|\boldsymbol{h}_{u}^{H}\boldsymbol{w}_{u}|^{2}}{\sigma^2})}{N},
\end{equation}
where $P$ is the power available per channel. For NOMA, since one cluster is formed per channel, $P$ represents the power available per cluster as we describe later in this section. 

\textcolor{black}{To model the user decoding capability constraints, we consider that each user-$u$ is associated with a decoding capability constraint $d$, represented as $d_u$. To illustrate the decoding capability constraint, the right side of Fig.\ref{fig:systemModel} shows the distribution of $N$ users to be served by the BS. Using the cosine similarity metric, each user will find the $b$ beams it is best aligned with, forming the user-beam set, $B_u$, for the user. As Fig.\ref{fig:systemModel} then highlights, each user has its SIC decoding capability associated with it. For example, user-$1$ with SIC capability of 0 ($d_{1}=0$) indicates it needs to be either served as an OMA user in an orthogonal channel of its own or in a NOMA cluster as the weakest user where it is not required to decode any other users' signals. User-$4$ with SIC decoding capability of 4 ($d_{4}=4$) indicates it is capable of decoding four other users' signals. This means that if user-$4$ is scheduled in cluster-$k$ at position $j$, then the maximum value of $j$ is 5 for this user since that would involve decoding 4 other users' signals, i.e., $\textnormal{max}(j) = 5$.}

\textcolor{black}{Let $d_{\textnormal{max}} = \textnormal{max}(d_u), \forall u=[1,..,N]$, represent the maximum decoding capability among the $N$ users in the system. If all users have the same decoding capability, i.e., $d_u = d_{\textnormal{max}}, \forall u=[1,..,N]$, we refer to this as homogeneous user decoding capabilities, or just a homogeneous system for short. In a homogeneous system, since any user-$u$ has the same decoding capability $d_u=d_{\textnormal{max}}$, this is equivalent to designing a user clustering scheme such that there are a maximum of $d_{\textnormal{max}}$ users per cluster. On the other hand, in a heterogeneous system, user clustering must be done in tandem with user ordering, such that each user in the cluster needs to decode at most $d$ other user's signals, where each user has its own value of $d$, $1 \leq d \leq d_\textnormal{max}$. This means that for every user-$u$ with decoding capability $d_u$ at SIC decoding position $j$ in cluster $C_k$, i.e., $\pi_k(j)$, it must satisfy that $d \geq j-1$. Using our nomenclature, $d_{\pi_k(j)}$ denotes the decoding capability of the $j$-th user in the $k$-th cluster. }

In this paper, the objective is to utilize NOMA to maximize the effective sum-rate of the system, such that each user's QoS is met and all user decoding capability constraints are satisfied. Let \textcolor{black}{$\Gamma_\textnormal{min}$} denote the minimum SINR with which each user needs to be served, \textcolor{black}{i.e., $\Gamma_{\pi_k(u)}^{\pi_k(u)} \geq \Gamma_\textnormal{min}$, $\forall u=[1,..,N]$}. The overall objective function to maximize $R_\textnormal{sum}$ can be stated as

\begin{subequations}
\begin{alignat}{2} 
&\underset{\textcolor{black}{\{C_k\}, \{w_{b_k}\},\{\pi_k\}, \{p_u\}}}{\text{max}} \qquad R_\textnormal{sum},\label{eq:overallObjective}\\
&\qquad \quad \ \text{s.t.} \quad R_u \geq \log_{2}(1+\Gamma_\textnormal{min}), \ \forall u = 1,..,N
\label{eq:constrainta1}\\
& \qquad \qquad \quad \ \ d_{\pi_k(j)} \geq j-1, \ j=\{1,..,\textcolor{black}{N_{C_k}}\}, \ \forall k=1,..,K, \label{eq:constraintb1}\\
& \qquad \qquad \quad \ \ \sum_{i=1}^{N_{C_k}} p_i \leq P, \ \forall k = 1,..,K,  \ \label{eq:constraintc1} 
\end{alignat}
\end{subequations}


\noindent where \eqref{eq:constrainta1} represents the QoS constraint, \eqref{eq:constraintb1} represents the decoding capability constraint, and \eqref{eq:constraintc1} represents the power per channel constraint.

In order to solve the optimization problem in \eqref{eq:overallObjective}, we break down the problem into two steps. First, we jointly tackle the user clustering, user ordering, and beamforming aspects, where we aim to minimize the number of clusters required to serve all the users while satisfying the beamforming and user decoding constraints. Second, once we have clusters of users, we do a power allocation step for the users in each cluster. We describe each of these steps next. 

In the first step, the goal is two-fold: a) to build clusters of SIC ordered users that satisfy the SIC decoding constraints and b) to identify which beam each of these clusters will be served by, such that the selected beam is in the user-beam set of each of the users selected to be in the cluster. The objective in this step is to serve all the users in the minimum number of clusters, while respecting the aforementioned constraints. Since each cluster is served on one orthogonal channel, the $N$ users being served on $K$ clusters, is equivalent to requiring $K$ orthogonal channel uses to serve the $N$ users. Hence, reducing $K$ improves the channel re-use, and in doing so, in general, contributes to an increased spectral efficiency. This is illustrated in equation \eqref{eq: Sum rate}, where $R_{\textnormal{sum}}$ is inversely proportional to the number of clusters, $K$. However, $R_{\textnormal{sum}}$ also depends on the SINR for each user in \eqref{eq: Sum rate}. This SINR for each user is affected by the other users they are clustered with, the order in which the users are decoded and finally the beamforming gain from the choice of beam from $B_c$ to serve each cluster with. Along with this, each users SIC decoding capability constraints need to be respected. Hence, we tackle these aspects jointly as a cluster minimization problem subject to several constraints as discussed in what follows next. 

For a single-cell NOMA deployment with no inter-cluster interference to consider since each cluster is served in an orthogonal channel, it is known that NOMA performance is significantly improved by decoding users in the order of their channel gains \cite{vaezi2019myths, ali2019clustering}. Hence, given that we have the full CSI of each user along with the precoding vectors, for every cluster $C_k$, we only allow the users to be decoded in the order of their effective channel gains. This means that the SIC decoding position-$j$ of user-$u$ with decoding capability $d_u$ in cluster-$k$, $\pi_k(j)$, is determined by the effective channel gain of the user-$u$ in relation to other users also selected to be in cluster $C_k$. Since we have the constraint that $d_{\pi_k(j)} \geq j-1$, we need to design clusters such that the users when ordered according to their effective channel gains, satisfy their SIC decoding constraints. Formally, the user clustering, user ordering, and BF optimization problem can be written as follows. Let $C = \{C_1, .., C_K\}$ represent the $K$ clusters required to serve the $N$ users. At most, each user is served in its own cluster or channel (equivalent to OMA), hence $K \leq N$ . Each cluster $C_k$ in $C$ represents a set of users ordered according to their effective channel gains when served by beam-$b_k$ from $B_c$ with precoding vector $\boldsymbol{w}_{b_k}$. Let $u_{i,k}$ be a binary variable that represents whether user-$i$ belongs to cluster-$C_{k}$, served by beam-$b_{k}$. Let $d_{\pi_k(j)}$ represents the user decoding capability of the $j^{\textnormal{th}}$ decoded user in cluster-$C_k$. The objective of our user clustering, ordering, and BF scheme is to minimize $K$, as follows:

\begin{subequations}
\begin{alignat}{2} 
&\underset{\textcolor{black}{\{u_{i,k}\}, \{b_k\},\{\pi_k\}}}{\text{min}} \qquad K,\label{eq:clusteringObjective}\\
&\qquad \quad \ \text{s.t.} \quad \sum_{k=1}^K u_{i,k}=1, \ \forall i = 1,..,N, \label{eq:constrainta} \\
& \qquad \qquad \quad \ \ b_k \in B_u, \ u_{i,k} =1, \ \forall k = 1,..,K, \ \label{eq:beamformingConstraint}\\
& \qquad \qquad \quad \ \ d_{\pi_k(j)} \geq j-1, \ j=\{1,..,\textcolor{black}{N_{C_k}}\}, \ \forall k=1,..,K,  \ \label{eq:decodingConstraint} \\
& \qquad \qquad \quad \ \ u_{i,k} \in \{0,1\}, \label{eq:constraintb} \\
& \qquad \qquad \quad \ \ b_k \in B_c,  \ \forall k = 1,..,K, \label{eq:constraintc} 
\end{alignat}
\end{subequations}


\noindent where constraint \eqref{eq:constrainta} ensures each user is placed in exactly one cluster. Constraint \eqref{eq:beamformingConstraint} is to ensure that the beam-$b_k$ chosen for cluster-$C_k$ in $C$ belongs to the user-beam list of each of the users selected to be served in that NOMA cluster. Constraint \eqref{eq:decodingConstraint} ensures the decoding capability constraints of each user in the system is adhered to. For the homogeneous system, since all users have the same decoding capability, we only need to limit the number of users per cluster. In other words, for a homogeneous system, within a cluster-$k$ of size \textcolor{black}{$N_{C_k}$}, if $N_{C_k} \leq d_{\textnormal{max}}$, then any decoding order within that cluster is feasible since all users have the same decoding capability of $d_u=d_{\textnormal{max}}$. Hence, for a homogeneous system, constraint \eqref{eq:decodingConstraint} can be simplified down to:

\begin{equation}
    \sum_{i=1}^N u_{i,k} \leq d_{\textnormal{max}}, \ \forall k=1,..,K.
\end{equation}


The second step is power allocation (PA), which is not the focus of this paper but briefly described here for completeness. Since only one cluster is served on one channel in our model, the channel power budget, $P$, is equivalent to the cluster power budget. Hence, the goal is to divide the power $P$, among the \textcolor{black}{$N_{C_k}$} users in each cluster $C_k \in C$. The objective in this step is to maximize the rate $R_k$ in each cluster. Since the users in the cluster are already ordered based on their effective channel gains, we iterate through the first $j = \{1,..,N_{C_k} -1\}$ users in the cluster at position $\pi_k(j)$ and assign it as much power as it needs to satisfy $\Gamma_{\textnormal{min}}$ and ensure successful SIC decoding, based on \eqref{eq: SINR of weak at the strong}. The strongest user is assigned the remaining power. We assume $P$ is always sufficient to meet each user's QoS, including the remaining power left over for the strongest user. This is similar to the QoS-based PA schemes described in \cite{sayed2016PAstrategies}.

\section{Proposed Algorithm(s)}\label{sec:algo}


\begin{algorithm}[!b]
\SetAlgoLined
\caption{NOMA-MEC} \label{Algorithm: NOMA-MEC}
\KwIn{Beam-list $B_u$ of $b$ beams for user $u$ with channel $\boldsymbol{h}_u$ and decoding capability $d_u$, $\forall u=[1,..,N]$. Also, precoding vectors of candidate beams, $\boldsymbol{w}_b$, $\forall b=[1,..,B]$.}
\KwOut{$K$ clusters of ordered users such that each user-$u$, is served in cluster-$k$ (with beam $b_k$) at position-$j$, such that $d_{\pi_k(j)} \geq j-1$ and $b_k \in B_u$, $k=[1,..,K]$, $\forall u=[1,..,N]$}

\textbf{Step-1}: Build candidate list $C_v$\;
\For{(\textnormal{beam}-$b$ : $B_c$)}{
Find all $n_b$ users that have beam-$b$ in $B_u$\;
Form set $C_t$ by computing all possible $\binom{n_b}{l}$ combinations, $\forall l=[2,..,d_{\textnormal{max}}]$\;

\For{($c$: $C_t$)}{
Order the users in $c$ according to the effective channel gains, creating set of ordered users in $c$ as $\{u_1,.., u_n\}$, such that $|\boldsymbol{w}_b^H \boldsymbol{h}_1|^2 \leq... \leq |\boldsymbol{w}_b^H \boldsymbol{h}_n|^2$

\For{$u_1$ to $u_n$}{
    \If{$d_{\pi_c(j)} < j-1$}{ 
    combination $c$ is invalid, skip it.
    
    break; (outer for loop)}
}
Add $c$ to candidate list $C_v$. (If we did not break, combination $c$ satisfies $d \geq j-1$ for all users in $c$).

}
}
Add each users- $u$, as a cluster of one with their best beam from $B_u$ to candidate list $C$, $\forall u=[1,..,N]$\;

\textbf{Step-2}: Run greedy MEC on $C_v$ to obtain $C$\;
$C_\textnormal{sorted}$ = sort $C_v$ in descending order\;
$x_c=0$ $\forall c \in C_v$\;
$C$ = \{\}, $K=0$\;
\For{$c_s$: $C_{\textnormal{sorted}}$}
{
\If{$z_{u,c}x_c \neq 0, \forall c \in C$ }
{
$C = \{C, c_s\}$\;
$K = K + 1$\;
}
}

return $K, C$;
\end{algorithm}


\begin{algorithm}[!t] 
\SetAlgoLined
\caption{NOMA-BB} \label{Algorithm: NOMA-BB}
\KwIn{\textcolor{black}{Best beam $B_u^*$} for user $u$ with channel $\boldsymbol{h}_u$ $\forall u=[1,..,N]$, common decoding capability $d_{\textnormal{max}}$ for all users, precoding vectors $\boldsymbol{w}_b$, $\forall b=[1,..,B]$.}
\KwOut{$K$ clusters of ordered users such that each user-$u$, is placed in a cluster served by its best beam, \textcolor{black}{$B_u^*$ and $N_{C_k} \leq d_{\textnormal{max}}$}, $\forall k=[1,..K]$.}

$C$ = \{\}, $K=0$\;
$m=d_{\textnormal{max}}$\;
\For{(\textnormal{beam}-$b$ : $B_c$)}{
Find all $n_b$ users that have \textcolor{black}{$B_u^* = b$} and group them in set $c:\{u_1,..,u_{n_b}\}$\;
\If{$n_b>m$}
{
\For{$i=1:\ceil[\big]{\frac{n_b}{m}}$}{
$j = \textnormal{min}(m \times (i+1), n_b)$\;
$C = \{C, \{u_{m \times i}, .., u_{j}\}\}$\;
$K = K + 1$\;
}
}
\Else{$C = \{C, c\}$\; 
$K = K+1$\;}

}

return $K$, $C$;
\end{algorithm}


In this section, we outline our two proposed algorithms, namely, the NOMA-MEC algorithm for heterogeneous systems in Algorithm \ref{Algorithm: NOMA-MEC} and the NOMA-BB algorithm for homogeneous systems in Algorithm \ref{Algorithm: NOMA-BB}.

We begin with the NOMA-MEC algorithm to solve the cluster minimization problem in \eqref{eq:clusteringObjective} for heterogeneous systems in Algorithm \ref{Algorithm: NOMA-MEC}. The goal is to minimize the number of clusters used while respecting the beamforming and user decoding capability constraints of each user in each cluster, as captured in \eqref{eq:beamformingConstraint} and \eqref{eq:decodingConstraint}, respectively. To do this, we break down the NOMA-MEC into two steps. In step-1, we find all possible valid cluster combinations, $C_v$, that respect both the constraints, \eqref{eq:beamformingConstraint} and \eqref{eq:decodingConstraint}. We refer to $C_v$, which is a set of valid user combinations, as the candidate list of clusters. Then, in step-2, from $C_v$, we find the minimum number of clusters that cover all the users exactly once. This is a MEC problem \cite{MEC}, hence we term the algorithm NOMA-MEC.

Step-1 of NOMA-MEC begins by building a list of users that can potentially be served on a NOMA cluster by each of the $B+1$ candidate beams in $B_c$. This is obtained by iterating through the user-beam set, $B_u$, of all users, $u = \{1,.., N\}$. Through this step, we get a list of users that can potentially cluster with each other. Let $n_b$ represent the number of users that have beam-$b$ in their user-beam set. Clusters can be of size $l = \{2,.., d_{\textnormal{max}}\}$. We treat clusters of one separately as described later in this section. Hence, we form all $\binom{n_b}{l}$ groups of users, for all $B+1$ beams. These are all potential clusters to be served in an orthogonal channel with a beam using precoding vector, $\boldsymbol{w}_b$. Along with $\boldsymbol{w}_b$, each user's channel vectors are known. Thus, we can order the users according to their effective channel gains from smallest to largest in each of these potential clusters. Since we only allow users to be decoded in this order, if any cluster has a user at position $\pi(j)$, such that $d_{\pi(j)}<j-1$, that cluster is invalid. Only those clusters that satisfy the decoding capability constraint \eqref{eq:decodingConstraint} for all the users in the cluster are added to the candidate list, $C_v$. Finally, all users can be in a cluster of their own and be served like they would be with OMA. Hence, we add $N$ elements to $C$, each being a cluster of one where user-$u$ is served on its best beam from $B_u$, $u=\{1,..,N\}$. 

In step-2 of NOMA-MEC, from the list of viable candidate cluster options in $C_v$, we want to select the minimum number of elements that would cover every user exactly once. Since we added clusters of one for each user in the last part of step-1, we are guaranteed the existence of a solution. Let $x_c$ be a binary variable that represents if element-$c$ from set $C_v$ is selected and $z_{u,c}$ be a binary variable that represents if user-$u$ belongs to element-$c$ in $C_v$. The optimization problem can be stated as follows: 

\begin{subequations}
\begin{alignat}{3} 
&\underset{x_c}{\text{min}} &\qquad& \sum_{c:C_v} x_c,\label{eq:objective5}\\
&\text{s.t.} &  & \sum_{u=1}^N z_{u,c}x_c =1, \ \forall c:C_v \label{eq:constraintb-5}\\
&  &  &  x_c, z_{u,c} \in [0,1], \label{eq:constraintc-5} 
\end{alignat}
\end{subequations}

\noindent where (\ref{eq:objective5}) represents the objective of the problem that minimizes the number of clusters and constraint (\ref{eq:constraintb-5}) ensures that all users occur exactly once in the final cluster set. This is a minimum exact cover problem, a known NP-complete problem \cite{MEC}. We solve this problem using a greedy algorithm as follows. The first step is to sort the clusters in $C_v$ in descending order of the number of users they contain, since using clusters in $C_v$ that cover the most number of users allows us to minimize the number of clusters we need to cover all the users. We then go through the list of cluster combinations, adding cluster-$c$ to $C$ only if all users in the cluster have not been covered by clusters already in $C$. The algorithm stops when all users have been covered exactly once, as highlighted in Algorithm \ref{Algorithm: NOMA-MEC}.

The complexity of the algorithm is influenced by the following parameters - 1) the number of beams each user picks in its beam set, $b$, which is an algorithm specific parameter, 2) the number of candidate beams, $B$, which is a system level design parameter that we can control and 3) the number of users, $N$, that need to be served along with their respective decoding capabilities, $d_u, u\in[1,N]$. In step-1 of NOMA-MEC, building $C_v$ involves the construction of $I_1$ clusters as follows:
\begin{equation}
    I_1 = \binom{n_b}{l}\times (B+1), \label{eq:I1}
\end{equation}
where $l=\{2,..,d_{\textnormal{max}}\}$. In each of these $I_1$ clusters, the users have to be ordered and then analyzed to check whether each users decoding capability criteria are staisfied. The parameter $n_b$, the number of users that have beam-$b$ in their user-beam set, scales with $N$ and $b$. The second step is the minimum exact cover problem. Let $I_2$ represent the number of valid combinations in $C_v$ that the greedy algorithm in MEC needs to explore to find $C$. In a homogeneous system, all the $Z$ clusters are valid clusters, which means $I_2=I_1$. Thus, a homogeneous system represents the worst-case complexity for the MEC part of the NOMA-MEC algorithm. However, in general, a large number of the original cluster combinations will be rejected due to them being unable to meet the user decoding capabilities, resulting in $I_2 << I_1$. This in turn controls the complexity of the MEC part of the algorithm. This is discussed further in Section \ref{sec:simulation}, supported by simulation figures.

The choice of $b$ is the most important design parameter for the NOMA-MEC algorithm. From a performance perspective, a larger $b$ gives the algorithm the ability to find a larger number of cluster combinations that satisfy the decoding capability constraint. So, strictly from the perspective of minimizing $K$ in \eqref{eq:clusteringObjective}, a large $b$ is good. However, as $b$ increases, we add beams that are less and less aligned with the user direction to the user-beam set $B_u$, reducing the beamforming gain with each increment of $b$. Due to the  overlapping nature of the beams as seen in Fig.\ref{fig:systemModel}, there is a value of $b$ such that the user is within the coverage area of all of its best $b$ beams in $B_u$. Let this value of $b$ be $b_{\textnormal{th}}$. However, as $b$ is further increased beyond $b_{\textnormal{th}}$, the user gets out of the coverage area of the beam-($b_{\textnormal{th}}$+$1$) and if NOMA-MEC schedules the user on beam-($b_{\textnormal{th}}$+$1$), it will have poor spectral efficiency, $R_u$, bringing down the overall spectral efficiency, $R_{\textnormal{sum}}$, in the process. The exact value of $b_{\textnormal{th}}$ depends on system-level parameters $M$ and $B$. The number of antennas, $M$, determines the width of the beam and together with the number of candidate beams, $B+1$, determines the level of overlap between the beams and hence also determines the value of $b_{\textnormal{th}}$. Additionally, $b$ is also an important parameter to control the complexity of NOMA-MEC. If the number of users is large or $d$ is large for most users, $b$ can be reduced to lower $I_1$. For a homogeneous system with a large $d_{\textnormal{max}}$, the number of combinations can be very large and so we need to scale back $b$. If $b=1$, it is equivalent to having each user pick its best beam. Hence, for the homogeneous system with large $d_{\textnormal{max}}$, we propose a low-complexity clustering algorithm called NOMA-best beam (NOMA-BB) that has each user served by its best beam in $B_u$, as we outline next.

In the NOMA-BB algorithm, like NOMA-MEC, we iterate through each beam and build the list of $n_b$ users that picked each beam-$b$. However, compared to NOMA-MEC, the difference  is that in this case, users have to belong to that beam since we follow the best beam strategy where each user picked only one beam in their user-beam set. Hence, these groups of users are effectively our clusters except that we might have beams that have more than $d_{\textnormal{max}}$ users in it, leading to some users needing to decode more than $d_{\textnormal{max}}$ users signals, which violates the SIC decoding capability constraint. Hence, for all beams where $n_b > d_{\textnormal{max}}$, we break up the one cluster of $n_b$ users into $\ceil[\big]{\frac{n_b}{m}}$\ clusters, where $m$ is an integer between 1 and $d_{\textnormal{max}}$, i.e., $m \in [1,d_{max}]$, that controls the maximum number of users per cluster and $\ceil{.}$ is the ceiling function. Since the goal is to minimize the number of clusters, we set $m=d_{\textnormal{max}}$. Setting $m=d_{\textnormal{max}}$ is feasible because all users have the same decoding capability, $d=d_{\textnormal{max}}$, and so any user ordering among the \textcolor{black}{$N_{C_k}$} users in some cluster-$C_k$ formed by NOMA-BB would be valid, as long as \textcolor{black}{$N_{C_k} \leq d_{\textnormal{max}}$}. In this paper, when we need to split $n_b$ users in a beam into multiple clusters, i.e., when $n_b > d_{\textnormal{max}}$, we arbitrarily split the users into different clusters. As a future work, a more advanced NOMA-BB clustering schemes could aim to maximize the channel disparity between the users in the cluster when doing this split, a condition known to improve the rate in NOMA systems \cite{higuchi2015non_chDisparity}. 


\section{Simulation Results and Discussion}\label{sec:simulation}

\begin{table}[!b]
\centering
\vspace{-1em}
\caption{Simulation Parameters}
\label{Tab: Simulation Parameters}
\begin{tabular}{|m{5.0cm}|m{2.55cm}|}
\hline
Parameter name, notation & Value \\ \hline
Number of paths, $L$ & $1$~\cite{cui2018unsupervised, cui2018OptimizationBased, Ding2017BF_mmWaveNOMA} \\
Path loss exponent, $\eta$ & $2$ \cite{cui2018unsupervised}\\ 
$ \frac{\textnormal{BS antenna spacing}}{\textnormal{Wavelength of the carrier signal}} $, $\frac{D}{\lambda}$ & $\frac{1}{2}$ \cite{cui2018unsupervised}\\ 
Noise power, $\sigma^2$ & $7.962 \times 10^{-11}$ \cite{cui2018unsupervised}\\
Users minimum QoS SINR, \textcolor{black}{$\Gamma_{\textnormal{min}}$}  & $0.02$ \cite{cui2018unsupervised}\\ 
Number of antennas at BS, $M$ & $[2,4,8,16,32]$\\
Number of candidate beams, $B$ & $[10,20,30,40]$ \\
Number of users in the system, $N$ & $[50,100,150,200]$ \\
Max. user decoding capability, $d_{\textnormal{max}}$ & $[0,10]$ \\ 
User distribution & Randomly distributed around $5$ meters radius from BS.~\cite{cui2018OptimizationBased}\\
\hline
\end{tabular}
\vspace{-1em}
\end{table}


The performance of the proposed NOMA-MEC and NOMA-BB algorithms are evaluated using MATLAB simulations, with the system parameters described in Table~\ref{Tab: Simulation Parameters}. The mmWave channel model in~\eqref{eq: channel response} is considered, where $L=1$, $\eta =2$ and $D/\lambda = 1/2$ for the ULA steering vector. The BS is equipped with $M=8$ antennas, unless specified otherwise. The noise power is $\sigma^2 = -174 + 10$log$_{10}(W) + N_f$ dBm, where $W=2$ GHz is the system bandwidth and the noise floor $N_f = 10$ dB. The users are randomly distributed around the BS within a 5 meter radius, i.e., $r_u \leq 5$. We consider the minimum user QoS to be an average of \textcolor{black}{$N \times \Gamma_{\textnormal{min}}$ bps per channel-use. Since the users are scheduled in $K \leq N$ channels ($K$ NOMA clusters), we can simplify this requirement by just considering a minimum user rate of $\Gamma_{\textnormal{min}}$ for each user in every cluster. In the simulations, $\Gamma_{\textnormal{min}} = 0.02$}. Finally, the number of candidate beams is $B=20$. In the simulations for heterogeneous systems, we set $d_{\textnormal{max}}=5$ and generate each user's decoding capability, $d$, as a random integer in the range $[0,d_{\textnormal{max}}]$. For the homogeneous systems, we vary $d_{\textnormal{max}}$ from 1 to 10.

\begin{figure}[!t]
    \centering
    \vspace*{-.02in}
    \subfloat[]{
        \hspace*{-.1in}
        \includegraphics[scale=0.55]{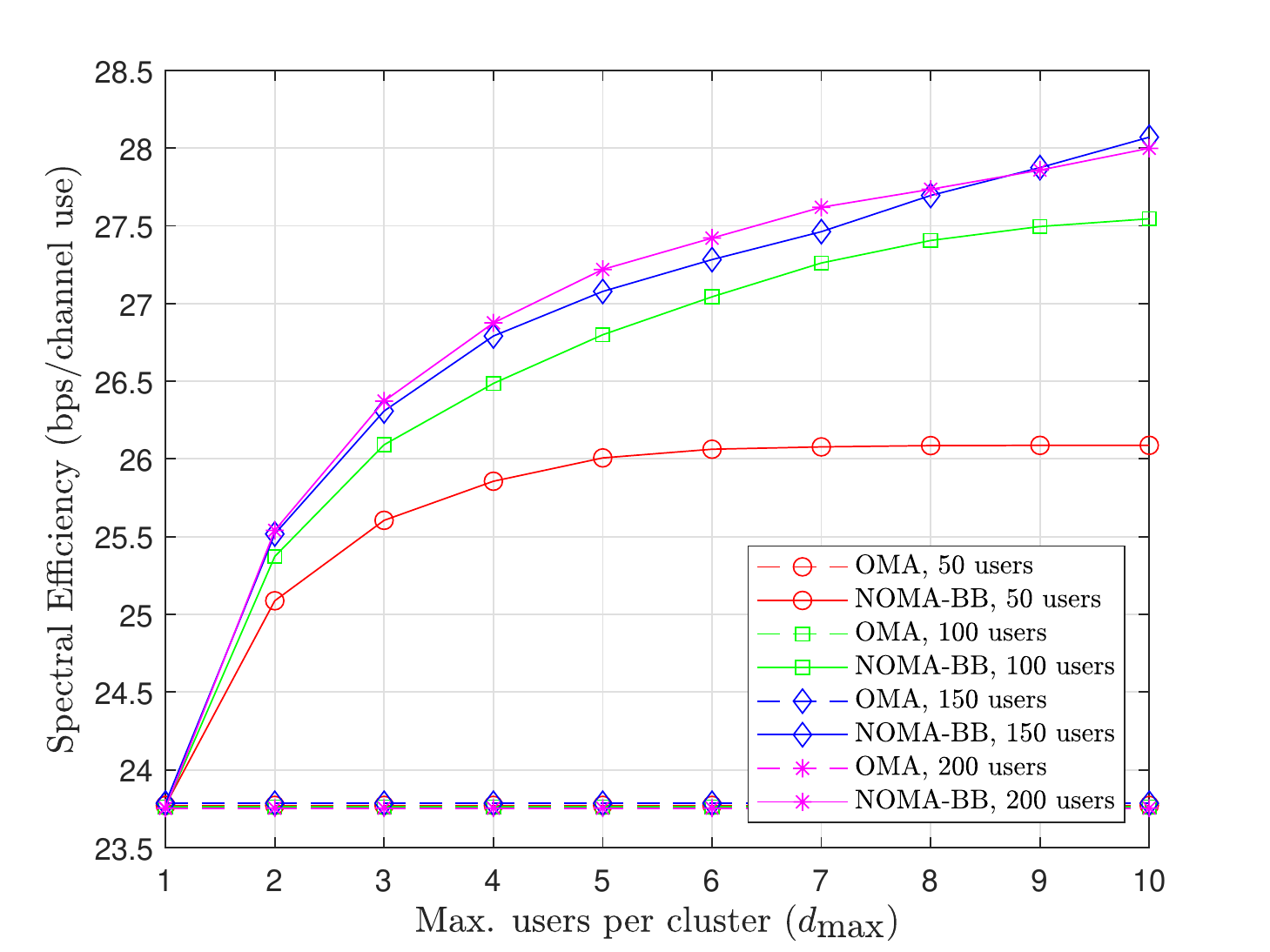}
        \label{fig:sim_NOMA-BB}
    }
    \vspace*{-.14in}
    \hfill
    \subfloat[]{
        \hspace*{-.1in}
        \includegraphics[scale=0.55]{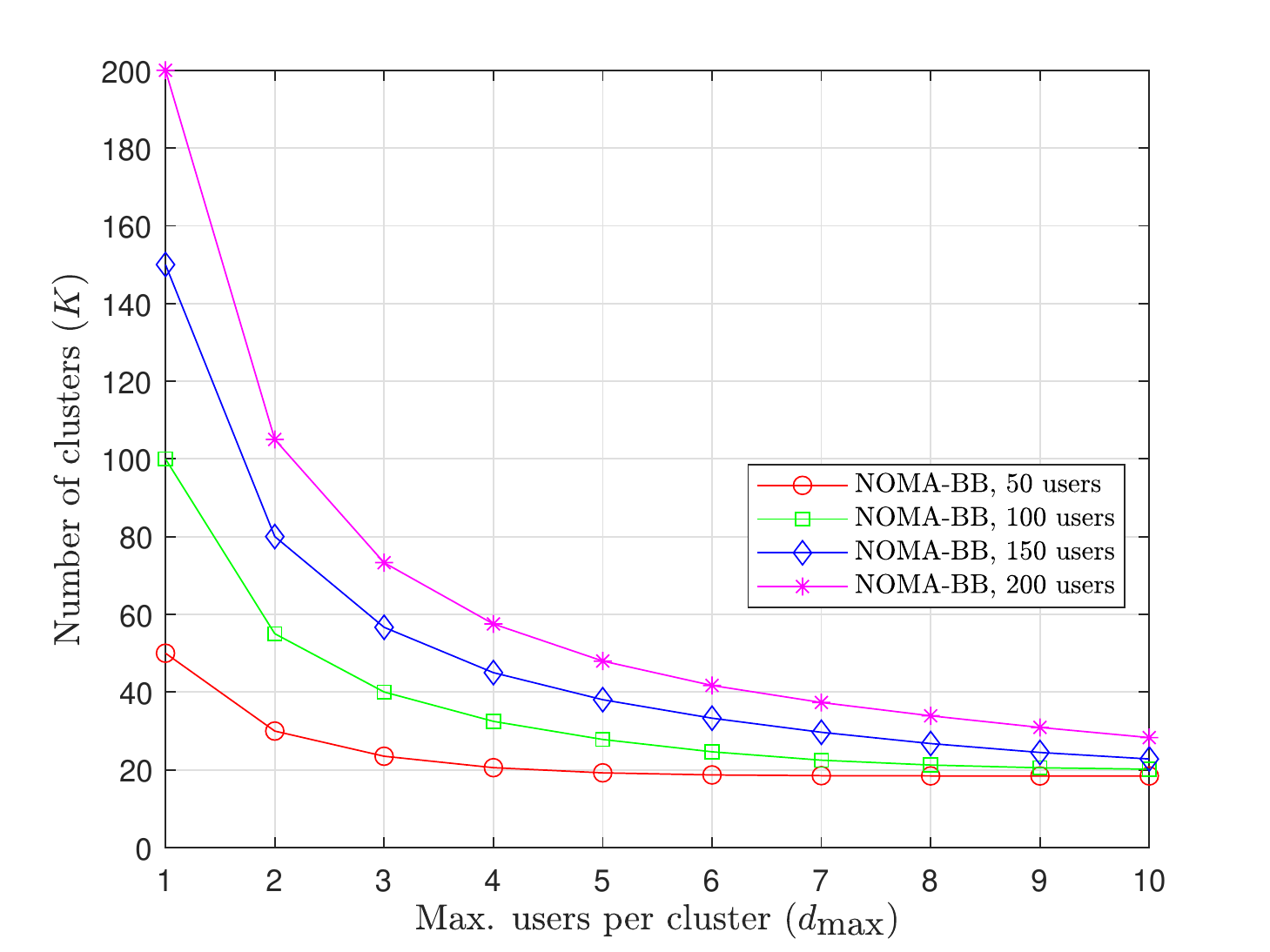}
        \label{fig:sim_NOMA-BB_clusters}
    }
    \caption{Performance of NOMA-BB for a homogeneous system where each users SIC decoding capability constraint $d=d_{\textnormal{max}}$, effectively making $d_{\textnormal{max}}$ the maximum number of users that can be placed in a cluster. (a) Overall spectral efficiency compared to OMA (b) Number of clusters, $K$, required to serve the $N$ users.}
    \vspace{-1em}
\end{figure}


We start by evaluating the NOMA-BB algorithm for a homogeneous system, where each users decoding capability is $d=d_{\textnormal{max}}$ that we vary from 1 to 10 in this simulation. In Fig.~\ref{fig:sim_NOMA-BB}, we compare the spectral efficiency, measured in \textcolor{black}{bps per channel use}, for the NOMA-BB algorithm against OMA for a system with 50, 100, 150, and 200 users. OMA is not influenced by the value of $d_{\textnormal{max}}$ as it has to serve one user per cluster, irrespective. As seen in Fig.~\ref{fig:sim_NOMA-BB}, a NOMA setting with $d_{\textnormal{max}}=1$ is equivalent to OMA. As $d_{\textnormal{max}}$ increases beyond one, we start to see the gain of NOMA. A higher value of $d_{\textnormal{max}}$ means all users are capable of decoding more number of other users signals, i.e., we can serve more users per-cluster. Looking at the NOMA-BB algorithm, for each beam-$b$, we split the $n_b$ users who picked beam-$b$ into $\ceil[\big]{\frac{n_b}{m}}$\ clusters, where $m=d_{\textnormal{max}}$. Clearly, as $d_{\textnormal{max}}$ increases, $m$ increases, and so NOMA-BB needs to form fewer clusters in this splitting step. This is illustrated in Fig.~\ref{fig:sim_NOMA-BB_clusters}, where the number of clusters, $K$ required to serve the $N$ users decreases as $d_{\textnormal{max}}$ increases. Further, as the number of users in the system, $N$, increases, the likelihood of having beams with more than $d_{\textnormal{max}}$ users increases in the first step of the NOMA-BB algorithm. As a result, for higher $N$, we see the number of clusters decrease in Fig.~\ref{fig:sim_NOMA-BB_clusters} by increasing $d_{\textnormal{max}}$ for longer before it starts to flatten out. Correspondingly, the rate in Fig.~\ref{fig:sim_NOMA-BB} increases with $d_{\textnormal{max}}$ for longer when $N$ is larger. 

\begin{figure}[t!]
\centering
\vspace{-1em}
\includegraphics[width=0.48\textwidth]{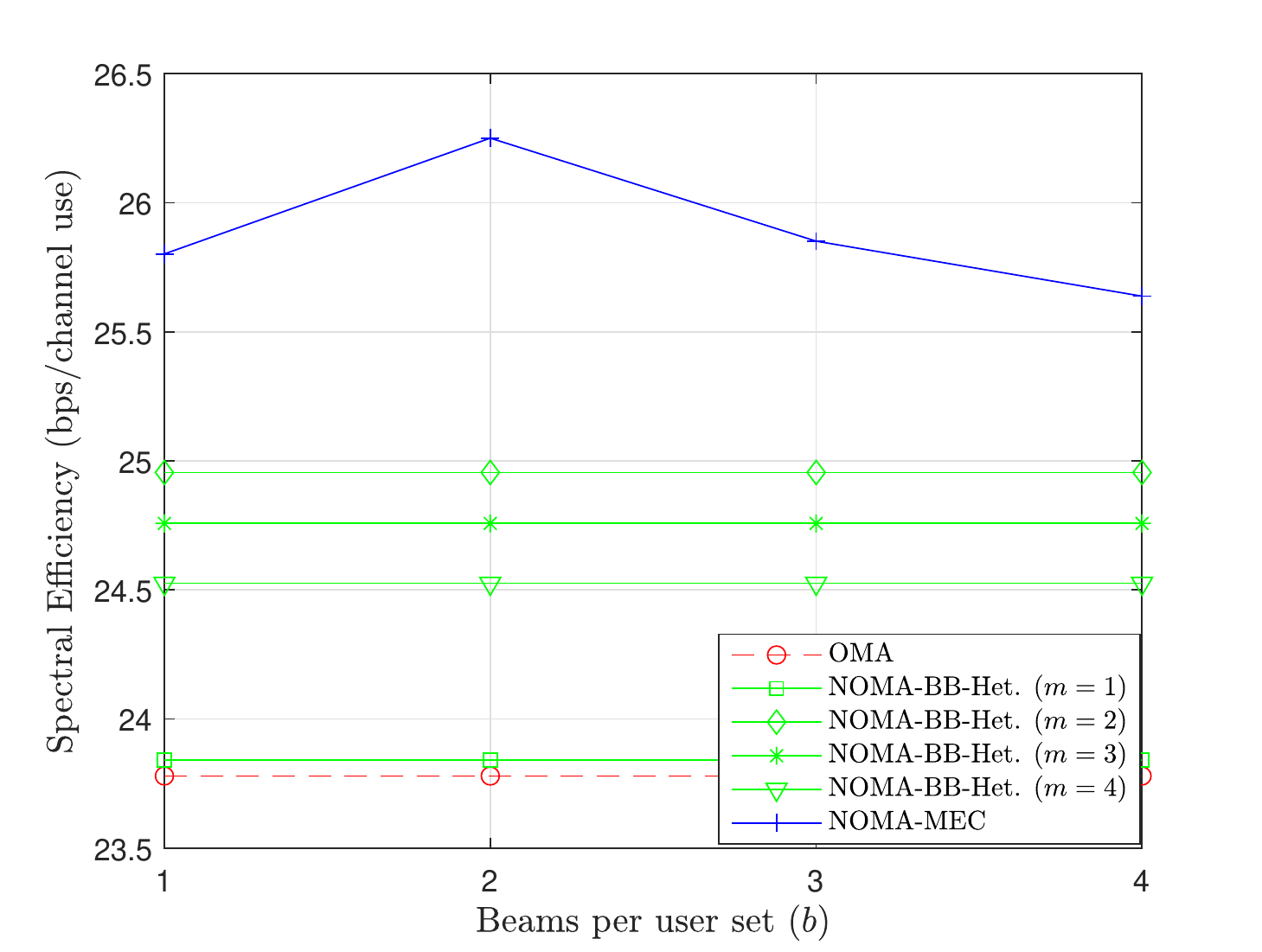} 
\caption{Performance of NOMA-MEC for a heterogeneous system where each user has its own decoding capability constraints. NOMA-MEC is compared against OMA and NOMA-BB-Het, a modified version of NOMA-BB to account for the heterogeneous user decoding capabilities.} 
\label{fig:sim_NOMA-MEC}
\end{figure}

We now move to heterogeneous systems and evaluate the performance of our proposed NOMA-MEC algorithm, compared against OMA and NOMA-BB with slight modifications to account for the heterogeneous decoding capability constraints. We note that there are no direct user clustering schemes in the literature that considers individual user decoding capabilities for us to compare against. NOMA-BB is fairly typical of most NOMA clustering schemes in the literature that do not have individual restrictions on each user's SIC decoding position, and so offers good insights for us to compare our proposed NOMA-MEC against. However, to run NOMA-BB in a heterogeneous system, we cannot set $m=d_{\textnormal{max}}$ like we could for a homogeneous system, since each user has its own decoding capability constraint and so not all clusters will result in feasible decoding order combinations, even if the cluster size is capped at $d_{\textnormal{max}}$. To make NOMA-BB work for a heterogeneous system, we need to separate out all users with $d<m$, for any $m \in [1,d_{\textnormal{max}})$, and then divide the remaining users into $\ceil[\big]{\frac{n}{m}}$\ clusters. This would ensure that the arbitrary user ordering done by the NOMA-BB scheme does not violate any user's SIC decoding capability constraint. A larger $m$ means we can form larger clusters but will have to exclude more users with the extreme case of $m = d_{\textnormal{max}}$ equivalent to OMA and so we exclude it, while a smaller $m$ means we will form smaller clusters, but exclude less users. We term this modified version of NOMA-BB for heterogeneous systems as NOMA-BB-Het and run it with all possible values of $m \in [1,d_{\textnormal{max}}), d_{\textnormal{max}}=5$, for the simulations in Fig.~\ref{fig:sim_NOMA-MEC} which we discuss next. 



\begin{figure}[t!]
\centering
\vspace{-1em}
\includegraphics[width=0.48\textwidth]{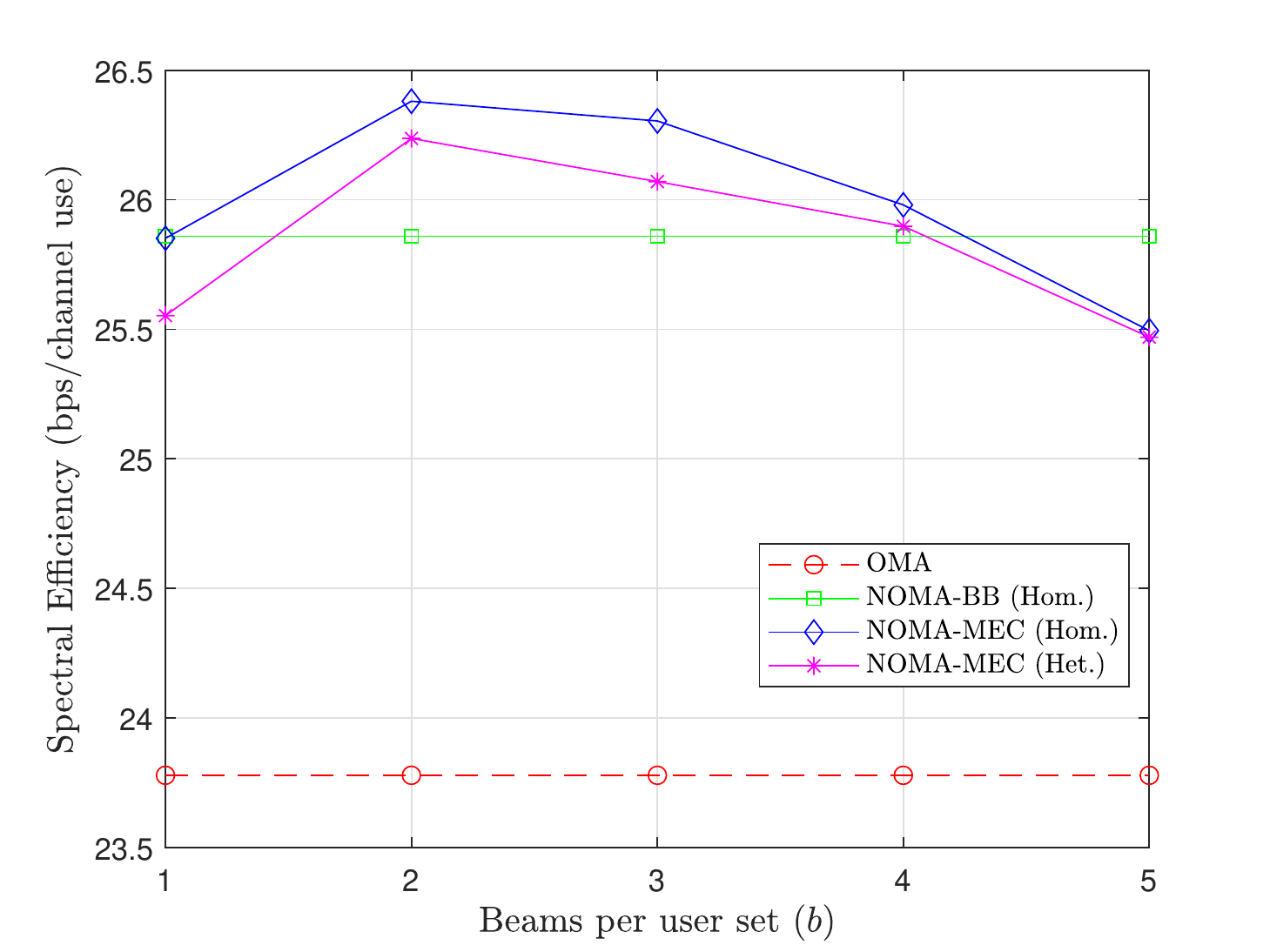} 
\caption{NOMA-MEC run in a heterogeneous system with users having their SIC decoding capability, $d$, randomly distributed in $[0,d_{\textnormal{max}}$], is compared against NOMA-BB and NOMA-MEC run for a homogeneous system where all users have $d=d_{\textnormal{max}}$ ($d_{\textnormal{max}}=5$).} 
\label{fig:sim_mixed}
\end{figure}

Analyzing the performance of NOMA-MEC from Fig.~\ref{fig:sim_NOMA-MEC}, we see that despite the restrictions put in place by the heterogeneous user decoding capabilities, we still see a significant performance gain over OMA. It also outperformed NOMA-BB-Het for all values of $m$, because NOMA-BB-Het and other such clustering algorithms from the literature do not consider restrictions on each individual user's capabilities while clustering. In Fig.~\ref{fig:sim_mixed}, the NOMA-MEC heterogeneous scheme running in a heterogeneous system with all users having a random value of $d$ in the range $[0,d_{\textnormal{max}}]$, is compared against a hypothetical homogeneous system where all users have $d=d_{\textnormal{max}}$. For the hypothetical homogeneous system, the original NOMA-BB and NOMA-MEC that assumes homogeneous user decoding capability with $m=d_{\textnormal{max}}$ is run. We note that the NOMA-MEC algorithm can easily be run for a homogeneous system, with all users having $d=d_{\textnormal{max}}$. We see that the NOMA-MEC algorithm for the heterogeneous deployment closely shadows, but always trails, the NOMA-MEC run for the homogeneous deployment. \textcolor{black}{The flexibility of the proposed NOMA-MEC algorithm is highlighted by this observation since it says that even though each user is posing its own decoding restrictions of $d \leq d_{\textnormal{max}}$, we are still able to achieve close to the performance we could if there was a simple maximum users per cluster constraint of $d=d_{\textnormal{max}}$. The hypothetical homogeneous deployment is still better because} in the NOMA-MEC for the homogeneous deployment, all $I_1$ cluster combinations examined is step-1 of NOMA-MEC are valid and entered into $C_v$ for the MEC algorithm to choose from. However, the NOMA-MEC for heterogeneous systems strips a large chunk of these $I_1$ combinations away due to not satisfying the user decoding capability constraints and so gives fewer pairing options in $C_v$ for the greedy MEC algorithm to work with when trying to minimize $K$. Looking at just the homogeneous curves in Fig.~\ref{fig:sim_mixed}, NOMA-MEC (Hom.) with $b \in [2,4]$ outperforms NOMA-BB. This is expected as the NOMA-MEC algorithm is more advanced, allowing users to pick multiple candidate beams for clustering, giving more clustering opportunities. .

\begin{figure}[t!]
\centering
\vspace{-1em}
\includegraphics[width=0.48\textwidth]{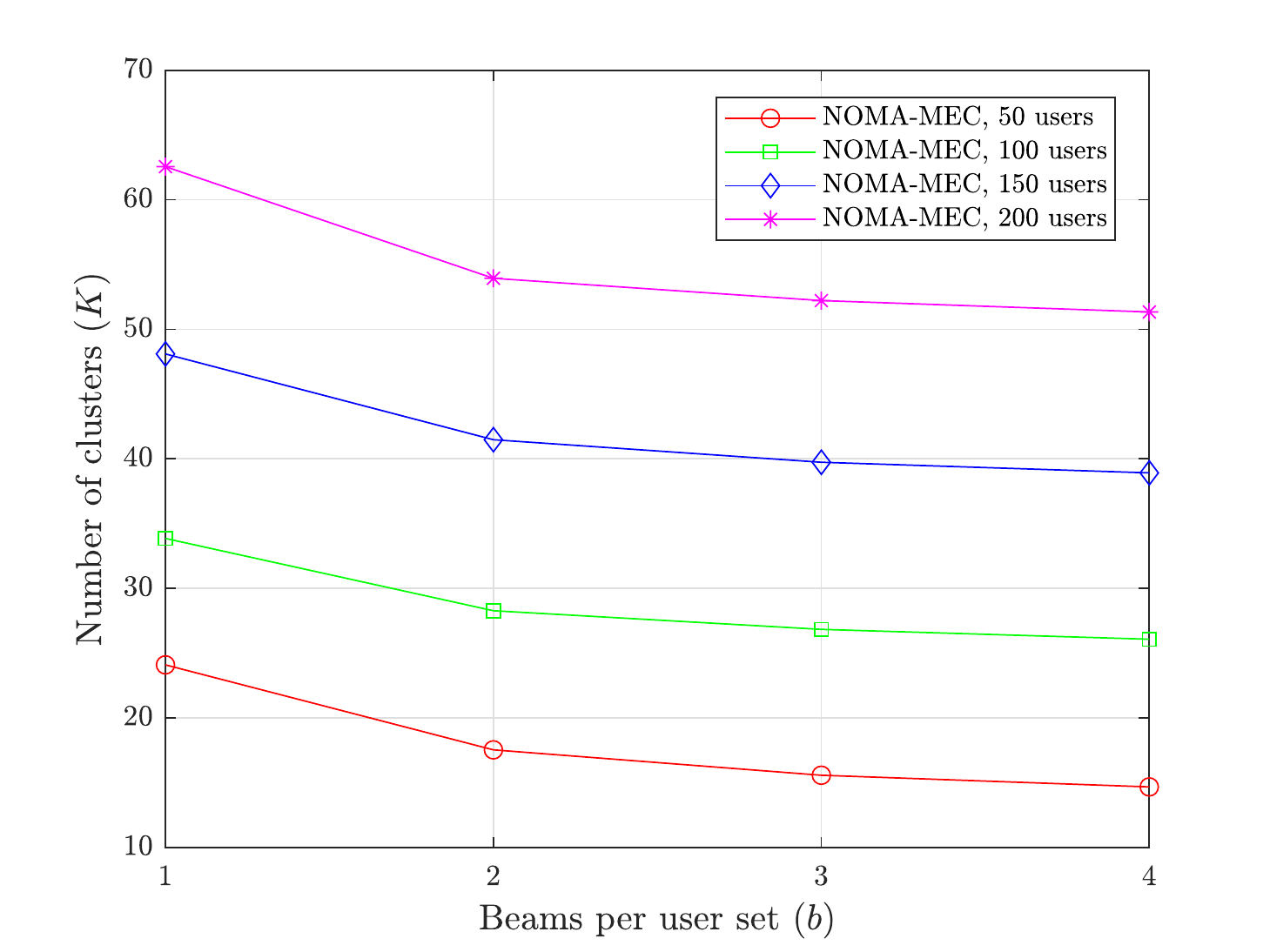} 
\caption{Analyzing NOMA-MEC in terms of the number of clusters, $K$, outputted by the algorithm, as the number of  heterogeneous users in the system increases for different values of $b$.} 
\label{fig:sim_NOMA-MEC_clusters}
\vspace{-1em}
\end{figure}


Additionally, analyzing the trends of NOMA-MEC in both Fig.~\ref{fig:sim_NOMA-MEC} and Fig.~\ref{fig:sim_mixed}, we see the rate increase at first as $b$ increases, but then starts to drop-off as we increase $b$ further. This is a consequence of the trade-off between a larger search space to reduce $K$ and the beamforming gain from allowing users to be served on their stronger beams, as discussed in Section~\ref{sec:algo}. A larger choice of $b$ implies a larger candidate cluster list $C_v$, in the NOMA-MEC algorithm, allowing the MEC part of the algorithm to find solutions with a lower number of clusters, $K$. This is illustrated in Fig.~\ref{fig:sim_NOMA-MEC_clusters}, where for any number of users in the system, the number of clusters required to serve the $N$ users, i.e., $K$, decreases as $b$ increases. However, as $b$ increases beyond the $b_{\textnormal{th}}$, users are adding beams to their user-beam-set, $B_u$, that they are less aligned with in terms of the cosine similarity metric. In other words, $\forall$ beam-$b$ in $B_u$, $b>b_{\textnormal{th}}$, NOMA-MEC can potentially schedule the user in a cluster served by beam-$b$, even though the user is out of the coverage area of beam-$b$. As seen in Fig.~\ref{fig:sim_NOMA-MEC} and Fig.~\ref{fig:sim_mixed}, at first as $b$ goes from one to two, the extra clustering opportunities allow us to reduce $K$ as well as not incur too much of a penalty in terms of the beamforming gain. However, as $b$ increases further, the penalty from sacrificing the beamforming gain outweighs the further cluster reduction we are able to achieve and hence we see the spectral efficiency start to drop off after that.  The exact value of $b$ at which this reversal occurs depends on the number of candidate beams, $B$ and the width of these candidate beams, which is a consequence of the number of transmit antennas, $M$. Next, we discuss the impact of $B$ and $M$ on the choice of $b$ from a performance perspective.

\begin{figure}[!t]
    \centering
    \vspace*{-.02in}
    \subfloat[]{
        \hspace*{-.1in}
        \includegraphics[scale=0.55]{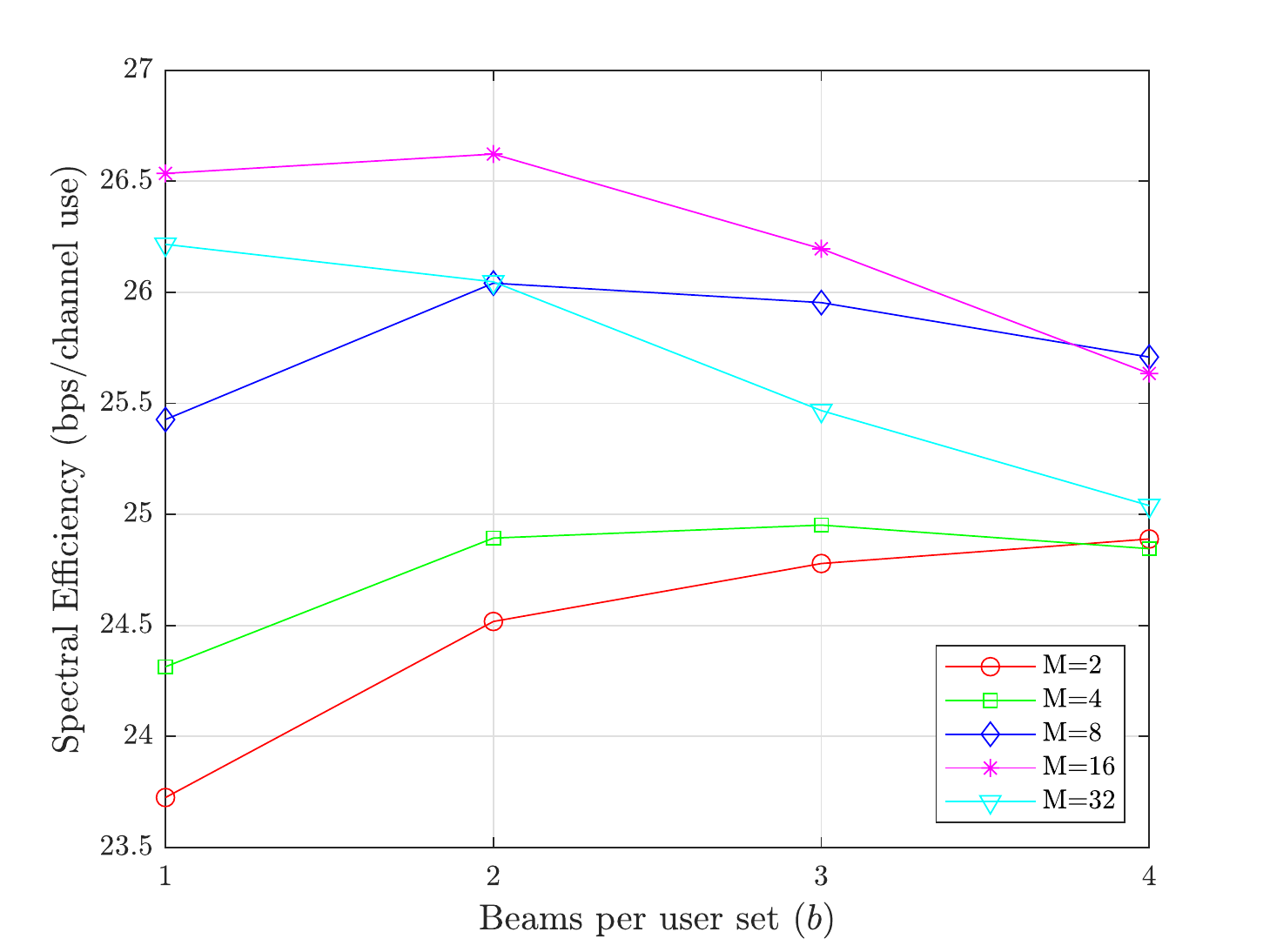}
        \label{fig:sim_varyM}
    }
    \vspace*{-.14in}
    \hfill
    \subfloat[]{
        \hspace*{-.1in}
        \includegraphics[scale=0.55]{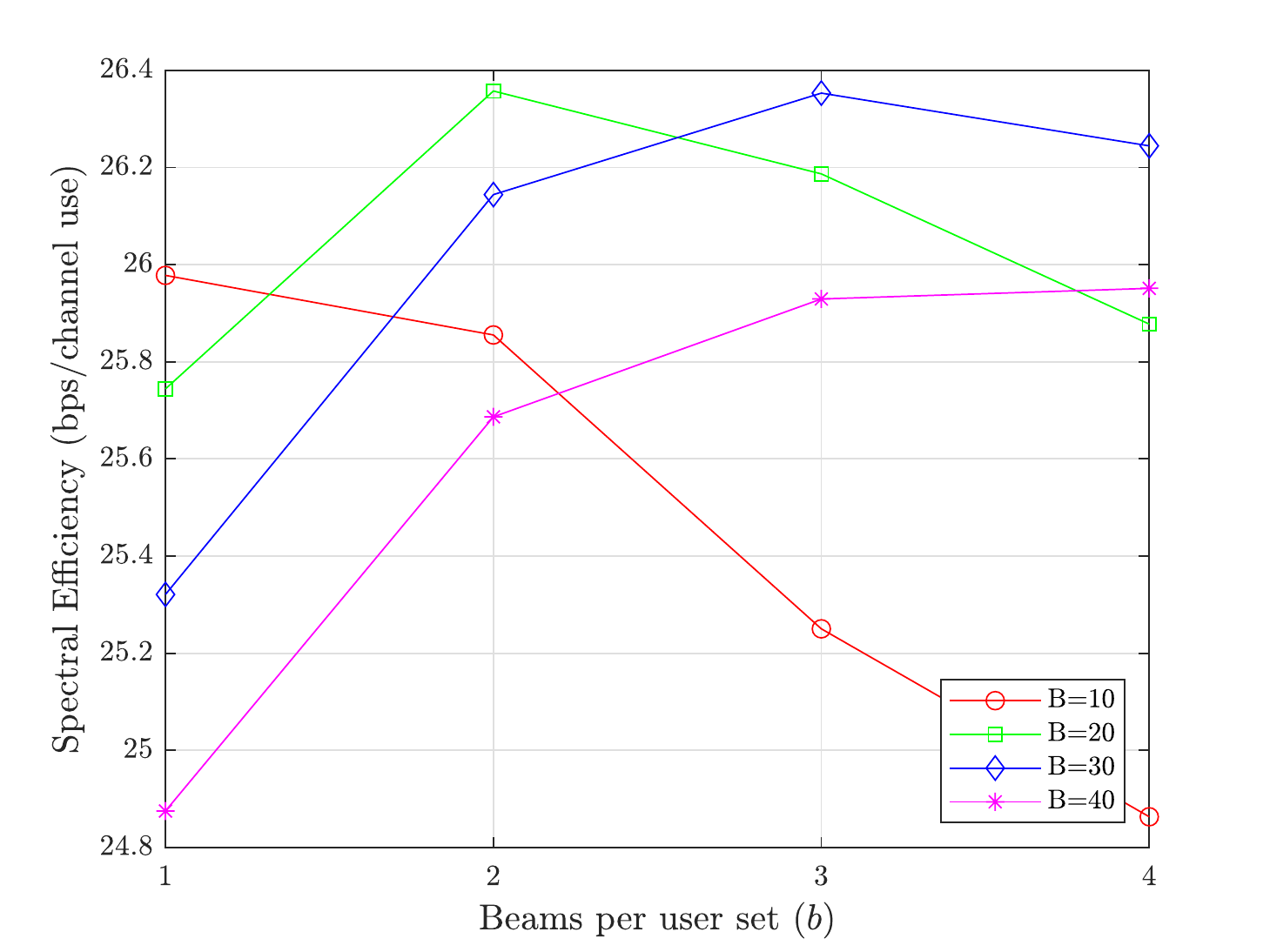}
        \label{fig:sim_varyB}
    }
    \caption{Analyzing the impact of $b$ on the performance of NOMA-MEC as system parameters $B$ and $M$ vary.}
\label{fig:sim_varyBM}
\vspace{-1em}
\end{figure}


The simulations in Fig.~\ref{fig:sim_varyBM} were conducted to understand the impact of parameters $B$ and $M$ respectively on the performance of NOMA-MEC. In particular, we are focused on the trend where we first see a performance improvement as $b$ increases, but then a decline as $b$ is further increased. In Fig.~\ref{fig:sim_varyM}, we vary $M$ while keeping $B$ fixed, $B = 20$. As the number of transmit antennas at the BS, $M$, increases, the BS is able to form more narrow beams. For a fixed value of $B$, as $M$ increases, the amount of overlap between the candidate beams decreases. This means that a user is located in the coverage area of a smaller number of beams, i.e., $b$ needs to be smaller to realize the beamforming gain. Conversely, as $M$ decreases for a fixed $B$, the overlap between the beams increases, as we have the same number of wider beams. This allows for a user to be located in the coverage area of more number of beams, i.e., a larger $b$ can be chosen. The same analysis can be done for a fixed $M$ and varying $B$ in Fig.~\ref{fig:sim_varyB}. In this case, the beam width is fixed since $M$ is fixed. However, as $B$ increases, the overlap increases as we have more candidate beams covering the same coverage area, $\bar\phi$. As a result, we see the performance gain from increasing $b$ for longer in Fig.~\ref{fig:sim_varyB} for larger values of $B$. We see that with every increase of $B=10$, the drop-off starts to occur $b=1$ later. Hence, irrespective of the values of $B$ and $M$, the trend is the same - we first see a gain from increasing $b$, but then the performance starts to drop off once we start to lose the beamforming gain as $b$ is further increased. For a small value of $M$ and a large value of $B$, the value of $b$ before the drop-off starts to occur is larger than if we had a small $B$ or large $M$. Hence, from a performance perspective, $b$ needs to be selectively tuned as a function of $B$ and $M$. We discuss the impact of parameter $b$ on the complexity of the algorithm next.

\begin{figure}[!t]
    \centering
    \vspace*{-.02in}
    \subfloat[]{
        \hspace*{-.1in}
        \includegraphics[scale=0.55]{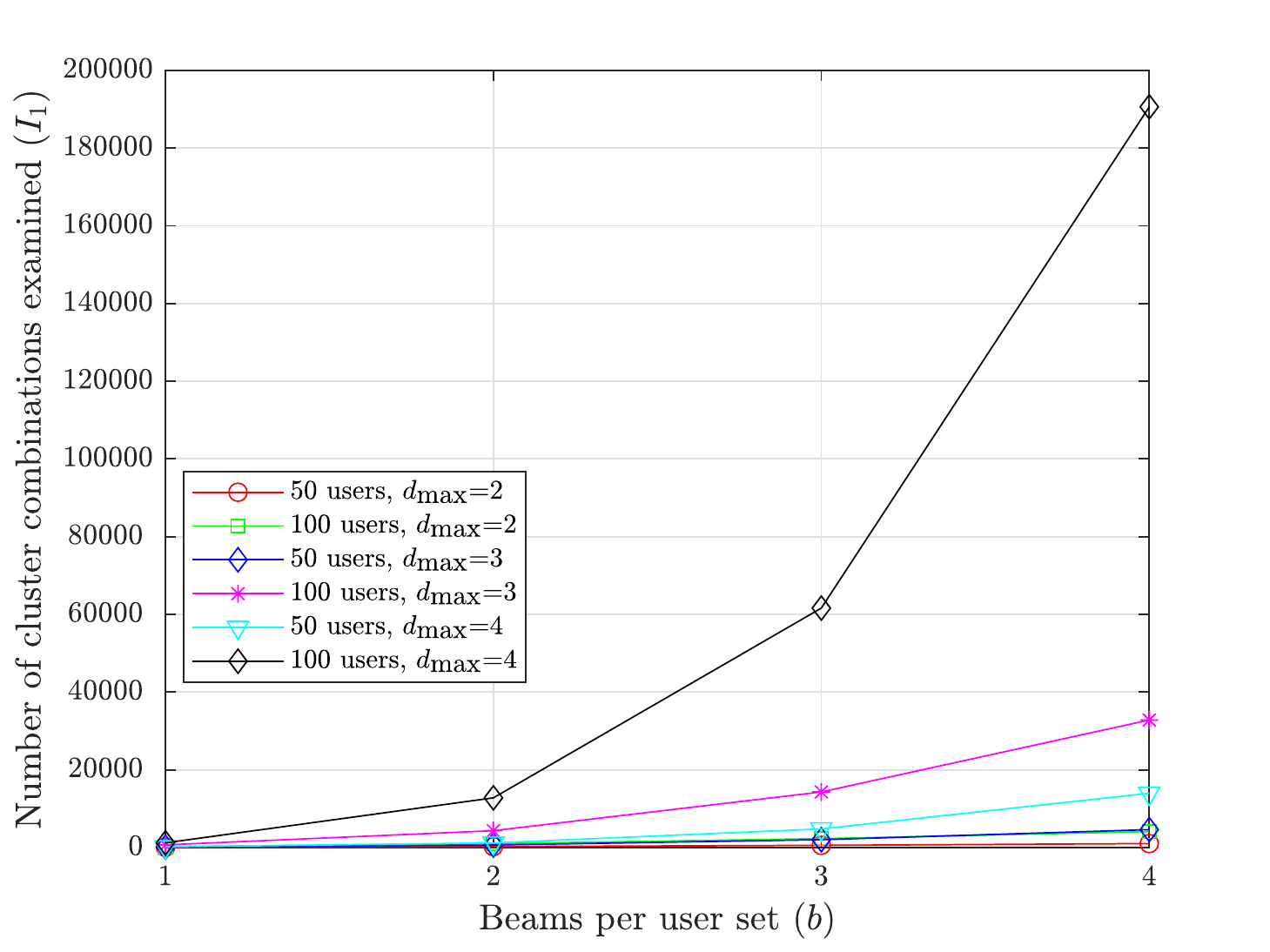}
        \label{fig:complexityA}
    }
    \vspace*{-.14in}
    \hfill
    \subfloat[]{
        \hspace*{-.1in}
        \includegraphics[scale=0.55]{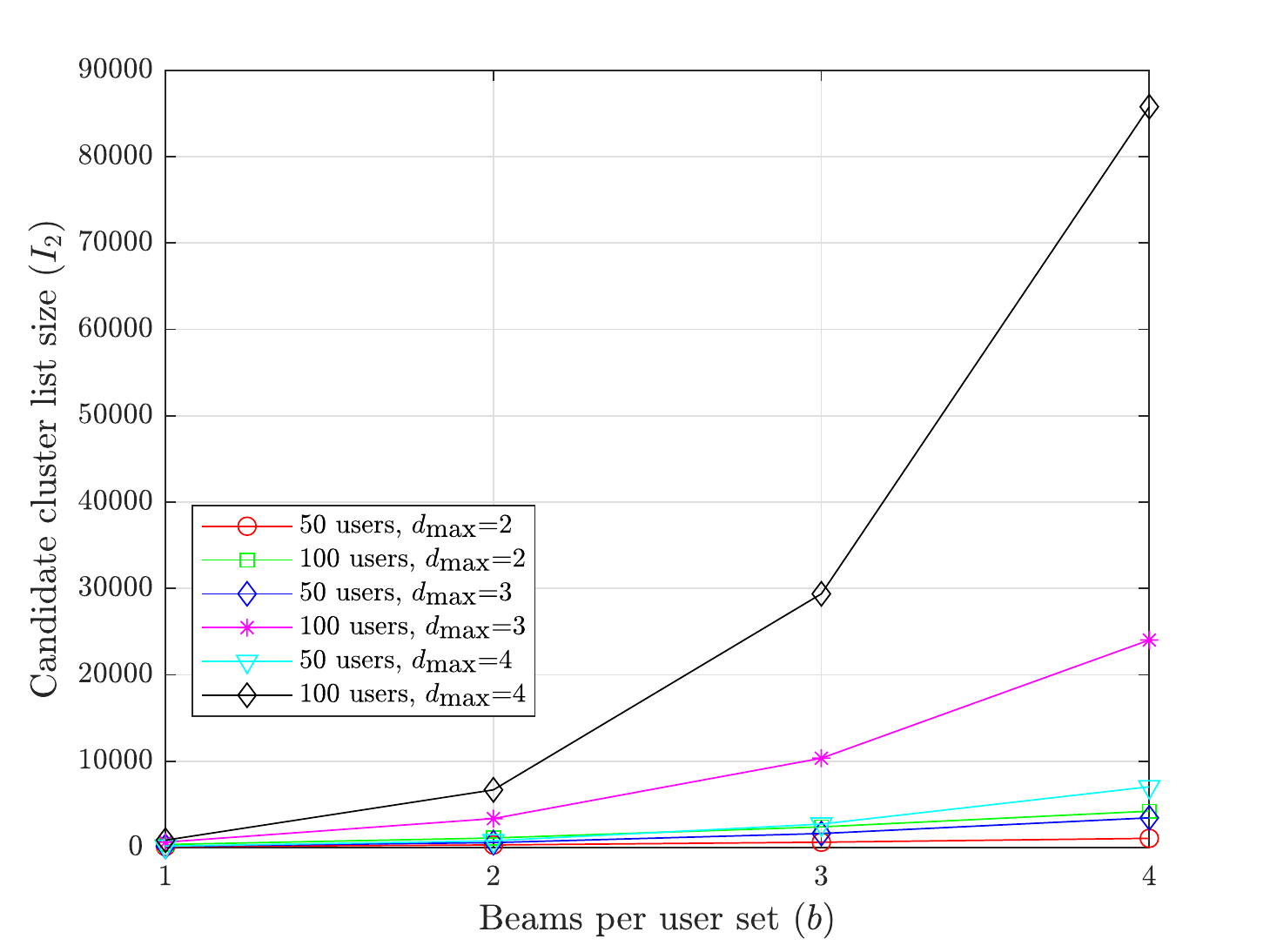}
        \label{fig:complexityB}
    }
    \caption{Impact of the number of beams per user set, $b$, on the complexity of NOMA-MEC in terms of (a) $I_1$ which represents the number of cluster combinations examined in step-1 of NOMA-MEC and (b) $I_2$, which represents the number of valid combinations in $C_v$ to be considered by the greedy algorithm to solve the MEC problem.}
\label{fig:complexityAB}
\vspace{-1em}
\end{figure}

As described in Section~\ref{sec:algo}, the NOMA-MEC algorithm presented in Algorithm~\ref{Algorithm: NOMA-MEC} consists of two steps. The first is the formation of candidate list $C_v$ by iterating through each beam in the list of candidate beams, looking for all valid combinations of users who could be scheduled together in a cluster served by the beam. Each possible combination requires users to be ordered by their effective channel gain and then the decoding capability of each user has to be checked to see if the combination is valid or not. The second part of the algorithm involves taking the candidate list, $C_v$, and running the greedy algorithm to solve the minimum exact cover problem. Simulation runs to present the number of iterations required at both steps of the NOMA-MEC algorithm, $I_1$ and $I_2$ respectively, are presented in Fig.~\ref{fig:complexityA} and Fig.~\ref{fig:complexityB} respectively.



The number of iterations in step-1 of NOMA-MEC, $I_1$, is determined by $n_b$, $l$ and $B$ as equation \eqref{eq:I1} shows. The term $n_b$, which corresponds to the number of users that contain each beam-$b$ in their user-beam-set, is influenced by the number of users in the system $N$ and the number of beams per user set, $b$. Since NOMA-MEC iterates through $\binom{n_b}{l}$, $l=\{2,..,d_{\textnormal{max}}\}$, combinations to check for valid cluster combinations based on the user decoding capability, the impact of $l$ on $I_1$ is entirely determined by $d_{\textnormal{max}}$.   The complexity of step-2 depends entirely on the size of set $C_v$, i.e., the number of valid combinations found after step-1, i.e., $I_2$. If $I_1$ is large, $I_2$ is likely to be large too and so the same factors that influence $I_1$ in step-1 also affect the complexity of step-2. However, since a large number of combinations examined in step-1 are deemed invalid and not added to $C_v$ in step-1 due to the SIC decoding capability restrictions, the size of $C_v$ will still be small. This is illustrated in Fig.~\ref{fig:complexityB}, where we see a significant reduction in the candidate cluster list size, $C_v$  compared to the number of cluster combinations examined  in Fig.~\ref{fig:complexityA}. As discussed in Section \ref{sec:algo}, if NOMA-MEC were run on a homogeneous system, $I_2 = I_1$, leading quickly to a prohibitively high complexity for the greedy algorithm to solve the MEC problem. On the other end of the spectrum, if a large number of combinations examined in step-1 are deemed invalid due to the SIC decoding capability constraints, the size of $C_v$ will still be small. This is likely in deployments where the majority of users are IoT users with $d \in [0,1]$ and there are only a handful of users with a larger value of $d$, e.g., cellular users. In such a case, the complexity of step-1 will be high since $d_{\textnormal{max}}=\textnormal{max}(d)$ is still large. However, if most users have $d\in [0,1]$, then most examined combinations in step-1 will be deemed invalid, still leaving a manageable size of $C_v$ for the greedy algorithm in step-2 to work with. However, in the simulations in Fig.~\ref{fig:complexityAB}, we only considered the case where the users' decoding capabilities, $d$, are randomly generated from $[0, d_{\textnormal{max}}]$. In future works, we will consider more skewed distributions of $d$ and explore how the complexity of step-1 of the NOMA-MEC algorithm can be reduced by exploiting advanced knowledge of the distribution of the users' decoding capabilities. 


Finally, it is worth mentioning that the parameter $b$ in the NOMA-MEC algorithm can be set by considering the performance-complexity tradeoff as follows. We have seen that increasing $b$ improves performance up to $b = b_{\textnormal{th}}$, but then the performance declines as $b$ is increased any further. Depending on system parameters $B$ and $M$, we can find the value of $b_{\textnormal{th}}$ at which the performance gains from increasing $b$ peaks. After that, the complexity aspect can be considered. If the complexity is acceptable at $b_{\textnormal{th}}$, that would be a logical choice for $b$. However, for systems with a large $N$ or $d_{\textnormal{max}}$, setting $b=b_{\textnormal{th}}$ could result in a prohibitively high algorithm complexity as seen in Fig.~\ref{fig:complexityAB}. In such settings, $b$ can be reduced to bring down the complexity as seen from Fig.~\ref{fig:complexityAB}, at the expense of performance.


\section{Conclusion}\label{sec:conclusion}

In this paper, we proposed a joint user clustering and user ordering scheme, namely, NOMA-MEC, for an ABF mmWave-NOMA system that can serve a set of users that each have their own SIC decoding capability constraints. By using the reported SIC decoding capability constraint from each user to set the maximum position in the SIC decoding order for clusters that are always decoded in the order of their effective channel gains, we framed the problem as a minimum exact cover optimization problem. Despite each user posing individual conditions on how many other users' signals it can decode in the SIC decoding order, simulation results demonstrated that the proposed algorithm still offers significant spectral efficiency gains over OMA as well as over other NOMA clustering algorithms that do not have the flexibility to accommodate for such user decoding requirements. We provided a detailed analysis of the performance-complexity trade-off from the setting of parameters related to the NOMA-MEC algorithm as well as system-level parameters. Finally, for a homogeneous system where all users have the same decoding capability requirements, we showed that this boils down to a simpler condition of restricting the number of users per cluster. We proposed a simpler NOMA-BB algorithm for the homogeneous system and also evaluated its performance through simulations.

\bibliographystyle{adityaIEEEtran}
\bibliography{IEEEabrv,main}

\vskip -2\baselineskip plus -1fil

\begin{IEEEbiography}[{\includegraphics[width=1in,height=1.25in,clip,keepaspectratio]{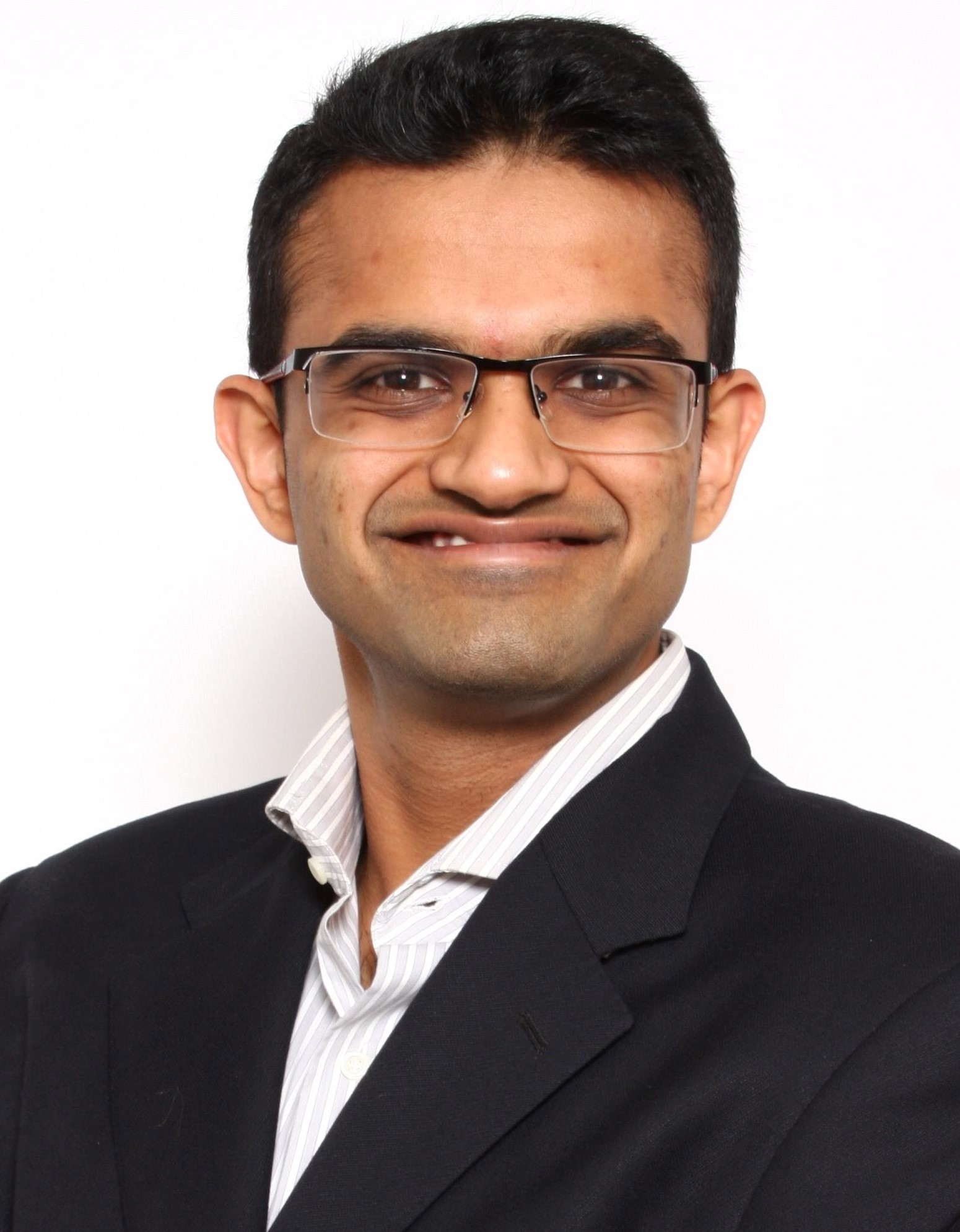}}]{\textbf{Aditya S. Rajasekaran}} (M'18) received the B.Eng (with High Distinction) and M.Eng degree in Systems and Computer Engineering from Carleton University, Ottawa, ON, Canada, in 2014 and 2017, respectively. He is currently pursuing his Ph.D. degree, also in Systems and Computer Engineering at Carleton University. His research interests include  wireless technology solutions aimed towards 5G and beyond cellular networks, including non-orthogonal multiple access solutions. 

He is also with Ericsson Canada, where he has been working as a software developer since 2014. He is currently involved in the physical layer development work for Ericsson's 5G New Radio (NR) solutions. 
\end{IEEEbiography}

\vskip -2\baselineskip plus -1fil

\begin{IEEEbiography}[{\includegraphics[width=1in,height=1.25in,clip,keepaspectratio]{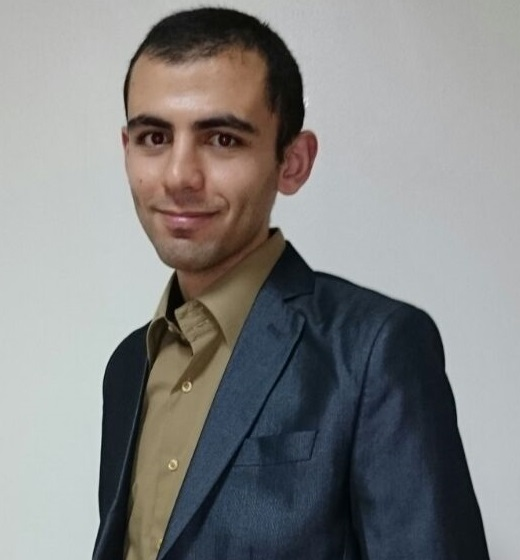}}]{\textbf{Omar Maraqa}}
has received his B.S. degree in Electrical Engineering from Palestine Polytechnic University, Palestine, in 2011, and his M.S. degree in Computer Engineering from King Fahd University of Petroleum \& Minerals (KFUPM), Dhahran, Saudi Arabia, in 2016. He is currently pursuing a Ph.D. degree in Electrical Engineering at KFUPM, Dhahran, Saudi Arabia. His research interests include performance analysis and optimization of wireless communications systems.
\end{IEEEbiography}

\vskip -2\baselineskip plus -1fil

\begin{IEEEbiography}[{\includegraphics[width=1in,height=1.25in,clip,keepaspectratio]{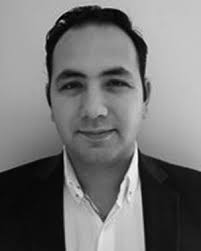}}]{\textbf{Hamza Umit Sokun}}
received the B.Sc. degree (Hons.) in electronics engineering from Kadir Has University, Istanbul, Turkey, in 2010, the M.Sc. degree in electrical engineering from Ozyegin University, Istanbul, Turkey, in 2012, and the Ph.D. degree in electrical engineering from Carleton University, Ottawa, ON, Canada, in 2017. Since 2017, he has been working for Ericsson, Ottawa, ON, Canada. His research interests include wireless communication and networks, signal processing, and machine learning.
\end{IEEEbiography}

\vskip -2\baselineskip plus -1fil

\begin{IEEEbiography}[{\includegraphics[width=1in,height=1.25in,clip,keepaspectratio]{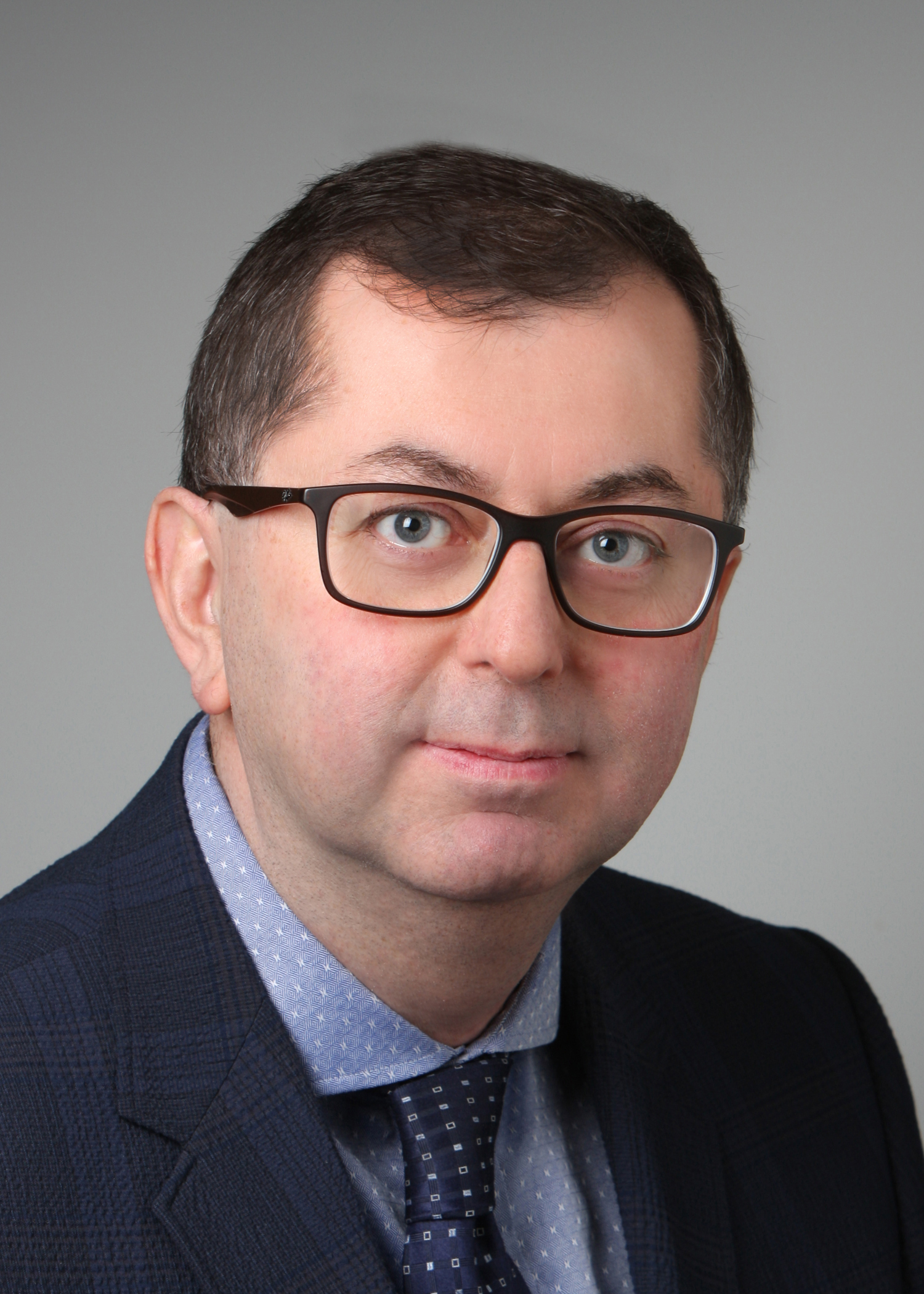}}]{\textbf{Halim Yanikomeroglu}} (F'17) is a Professor at Carleton University, Canada. His research covers many aspects of communications technologies with emphasis on wireless networks. He is a Fellow of IEEE, a Fellow of the Engineering Institute of Canada (EIC), a Fellow of the Canadian Academy of Engineering, a Distinguished Speaker for the IEEE Communications Society and the IEEE Vehicular Technology Society. He has been one of the most frequent tutorial presenters in the leading international IEEE conferences. He has had extensive collaboration with industry which resulted in 37 granted patents (plus more than a dozen applied). During 2012-2016, he led one of the largest academic-industrial collaborative research projects on pre-standards 5G wireless, sponsored by the Ontario Government and the industry. He served as the General Chair and Technical Program Chair of several major international IEEE conferences. He is currently serving as the Chair of the Steering Board of the IEEEs flagship wireless conference, WCNC (Wireless Communications and Networking Conference). He supervised 26 PhD students (all completed with theses).
\end{IEEEbiography}

\vskip -2\baselineskip plus -1fil

\begin{IEEEbiography}[{\includegraphics[width=1in,height=1.25in,clip,keepaspectratio]{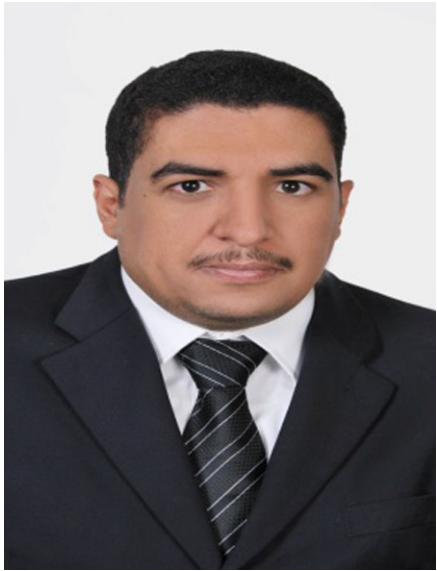}}]{\textbf{Saad Al-Ahmadi}}
has received his M.Sc. in Electrical Engineering from King Fahd University of Petroleum \& Minerals (KFUPM), Dhahran, Saudi Arabia, in 2002 and his Ph.D. in Electrical and Computer Engineering from Ottawa-Carleton Institute for ECE (OCIECE), Ottawa, Canada, in 2010. He is currently with the Department of Electrical Engineering at KFUPM as an Associate Professor. His current research interests include channel characterization, design, and performance analysis of wireless communications systems and networks.
\end{IEEEbiography}

\end{document}